\documentclass[aps,a4paper]{revtex4-2}
\usepackage{epsfig}
\usepackage{graphicx}
\usepackage{amsmath,amssymb,color}
\usepackage[english]{babel}
\usepackage{natbib}
\usepackage{xcolor}
\usepackage{float}
\parskip=\medskipamount
\usepackage{subfigure}

\newcommand{\eq}[1]{(\ref{#1})}
\newcommand{\fig}[1]{Fig.~\ref{#1}}

\newcommand{\be}{\begin{equation}}
\newcommand{\ee}{\end{equation}}
\newcommand{\beq}{\begin{equation}}
\newcommand{\eeq}{\end{equation}}
\newcommand\disp{\displaystyle}

\newcommand{\la}{\left<}
\newcommand{\ra}{\right>}

\begin{document}

\title{Topological localization in unentangled random braids}

\author{Daniil Fedotov$^{1,2}$, Pieter Hendrik Willem Van Der Hoek$^{3}$, and Sergei Nechaev$^{4,5}$}
\affiliation{$^1$UFR Sciences, Universit\'e Paris Saclay 91405 Orsay Cedex, France \\ $^2$IMT Atlantique 655 Av. du Technopôle, 29280 Plouzané, France \\ 
$^3$Molecular and Statistical Biophysics SISSA Via Bonomea 265 34136 Trieste, Italy \\ $^4$LPTMS, CNRS--Universit\'e Paris Saclay 91405 Orsay Cedex, France \\
$^5$Laboratory of Complex Networks Center for Neurophysics and Neuromorphic Technologies, Moscow, Russia}

\begin{abstract}

In 2015, P. Serna, G. Bunin, and A. Nahum \cite{serna} numerically investigated fluctuations in a bunch of unentangled random threads. They showed that fluctuations of an individual thread in a braid are qualitatively different from those of a thread embedded in a medium of fixed parallel lines. In particular, Serna et al. observed a logarithmic increase in the average transverse span, $\la|x_N|\ra$, of a thread with polymer length, $N$. This behavior highlights the collective nature of topological interactions, in which each polymer fluctuates and constrains the motion of the others. Studying this problem both numerically and analytically, we propose the following interpretation: the logarithmic regime is transient, and sufficiently long polymers exhibit topological localization, where $\la|x_N|\ra$ approaches a finite limit at $N\to\infty$. Unlike localization induced by quenched external disorder, this phenomenon arises solely from collective topological constraints. We study braids within a locally free approximation proposed in \cite{vershik2000} and support the localization by analytical investigation of Brownian bridges on the asymptotic Cayley graph of a locally free group. We further conjecture that topologically induced localization emerges as a phase transition: for a small number of threads, $n \sim 2\div 5$, the span, $\la|x_N|\ra$, grows algebraically, with the usual scaling $\la|x_N|\ra\sim N^{1/4}$ at $N\gg 1$; by contrast, for sufficiently large $n$, each thread in a long braid ($N\gg 1$) asymptotically localizes, approaching $\la|x_N|\ra\sim \mathrm{const}$.

\end{abstract}

\maketitle

\tableofcontents

\section{Introduction}
\label{sect:01}

Topology gives rise to interactions that are fundamentally different from conventional energetic ones. While energetic interactions are local, i.e. depend directly on interacting objects, topological constraints are global -- they restrict the space of accessible configurations \cite{Micheletti2011, Tubiana2024}. Such constraints are responsible for many remarkable properties of extended objects, like for example polymeric systems, ranging from the rheology of melts to chromosome organization and ring-polymer statistics \cite{FrankKamenetskii1975, Halverson2014, Rosa2014, sakaue2011}. Despite essential progress in understanding and description of topological interactions, yet the understanding collective statistical mechanics of topologically constrained many-body systems still remains a challenging task. Even in the absence of energetic interactions, topology alone can generate strong correlations extending over the entire system, and the mechanisms by which these correlations emerge are remain often out of a clear understanding.

Directed polymer fibers provide one of the cleanest realizations of a purely topological many-body problem. Consider an ensemble of $n$ directed polymer threads, each consisting of $N$ steps, connecting two parallel planes (from top to bottom). Such a configuration can be viewed as a "braid" \cite{Artin1947}. Since polymers cannot pass through one another, each configuration belongs to a particular isotopy class of braids. Two configurations are topologically equivalent if one can be continuously deformed into the other while keeping the endpoints fixed and forbidding strands from passing through one another. The statistical ensemble of braids is therefore naturally decomposed into sectors labeled by elements of the braid group \cite{Artin1947, Rocklin2012}.

The simplest, yet very conceptually rich, sector is the trivial braid, i.e., the braid isotopic to a collection of parallel straight threads. Geometrically, however, a trivial braid need not resemble straight lines. A large number of local strand exchanges may occur, provided they cancel globally so that the braid word reduces to the identity. Such random trivial braids represent highly disordered configurations constrained only by global topological triviality. They form a natural model of directed polymer melts in which all polymers remain collectively unentangled.

A central observable is the transverse wandering of an individual thread. Let $x^{(i)}_y$ denote the transverse position of the $i$-th polymer thread at longitudinal coordinate $y$ ($0<y<N$). The corresponding span, $|x^{(i)}_N|$ is defined as the maximal transverse excursion along the sample,
\be
|x_N^{(i)}|=\max_{y\in[0,N]}|x^{(i)}_0-x_y^{(i)}|,
\label{eq:01}
\ee
which defines the width of the thread's transversal fluctuations in the bundle. The scaling of the average span with polymer length, $N$, provides a direct measure of the degree of localization induced by topological constraints. Since all threads in the braid are identical, later we omit the index $i$ in $|x_N^{(i)}|$ considering a behavior of a "typical" thread denoted as $|x_N|$.

In 2016, P. Serna et al. \cite{serna} performed extensive numerical simulations of directed polymer melts restricted to the trivial braid sector. Their work showed that topological constraints alone dramatically suppress transverse fluctuations compared with those of unconstrained directed polymers. More importantly, they found that the fluctuations of an individual thread in a random trivial braid differ qualitatively from those of a single topologically trivial thread embedded in an array of fixed parallel lines. The latter corresponds to an individual topological constraint, whereas the former is an intrinsically collective problem in which every polymer fluctuates while simultaneously acting as a topological constraint on all the others. Their simulations therefore provided compelling evidence that collective topological interactions cannot be reduced to an effective external disorder generated by a regular array of threads treated as immobile obstacles. The origin and interpretation of this collective localization have remained open questions. In our view, \cite{serna} is a seminal yet underappreciated work that raised fundamental questions deserving careful investigation.

Based on numerical results, in \cite{serna} it has been supposed that the typical transverse span, $\la|x_N|\ra$ increases logarithmically with the polymer length, $N$, following the scaling law $\la|x_N|\ra\sim (\ln N)^{\zeta}$, where $\zeta$ depends on the model under consideration: $\zeta \approx 1.5$ for the full braid group, and $\zeta\approx 1.0$ for the locally-free approximation $LF_n$ of the braid group. Such a behavior already implies a substantial suppression of fluctuations compared with free directed polymers, yet it also suggests that localization is never complete, no matter how long the polymers is. Whether logarithmic growth represents the true asymptotic behavior or merely a long crossover has remained an unanswered question.

In our work we revisit this problem and arrive at a slightly different physical picture. We argue that the logarithmic regime is transient and that the true long-thread behavior at $N\to\infty$ is characterized by topological localization: $\la|x_N|\ra\to \mathrm{const}$ at $N\to\infty$. Thus, unlike localization induced by quenched disorder or external confinement, the topological localization emerges purely due to topological constraints.

The mechanism underlying this behavior is fundamentally collective. Every transverse displacement changes the topological accessibility of neighboring threads. For purely entropic reasons, arising exclusively from the structure of the braid group, fluctuations of threads near their straight configurations are entropically more favorable than wandering in the bulk. This leads to the spontaneous localization of unentangled threads. Since all threads are identical, this behavior results in emergent localization in the whole trivial braid: each thread fluctuates within a narrow strip around its straight configuration.

To demonstrate that fluctuations of topologically trivial threads are highly sensitive to the formation of entropic traps, and that the behavior of a test chain in a regular array of parallel threads is therefore unstable, we consider a model of a single unentangled chain in an effective \emph{inhomogeneous} lattice of parallel threads. By creating a hole in the regular array of lines, we introduce an "entropic trap" which, as we show, immediately induces the localization of a test chain at the hole.

Our results provide theoretical grounds for the phenomenon of “topological localization”, motivated by the observations of \cite{serna}, while revising their interpretation. A frozen \emph{homogeneous} lattice of topological obstacles imposes constraints on a fluctuating polymer, whose configurational properties are highly sensitive to the exact lattice symmetry. In contrast, a trivial braid constitutes a dynamically self-generated topological medium whose collective fluctuations directly feed back onto every constituent thread. It is precisely this feedback loop that drives asymptotic localization and gives rise to stable heterogeneous topological domains.

An immediate consequence of our approach is the existence of a collective topological phase transition as a function of the number of threads. For sufficiently small braid widths, $n$, an individual thread behaves essentially as if it were moving through a homogeneous array of obstacles, reproducing the asymptotic behavior of the single-polymer problem. Above a critical number of threads, however, the collective feedback generated by the fluctuating trivial braid qualitatively changes the dynamics, driving the system into a localized regime in which each thread becomes confined to a finite transverse domain. Thus, increasing the braid width, $n$, induces a transition from a "regular lattice of obstacles"--dominated regime to a collective topological phase characterized by self-induced topological localization.

More broadly, our work identifies the trivial braid as a nontrivial collective state of matter. Although its global topology is equivalent to that of parallel straight threads, the resulting localization transition arises as a purely collective phenomenon due to the "topological instability". We expect that this mechanism extends well beyond directed polymer bundles and provides a general mechanism for localization induced by topological interactions in many-body systems.

The topological localization may have important consequences for the crumpled globule model \cite{grosberg1988, grosberg1993, lieberman2009} of genome organization. An intriguing possibility is that localization could control the structure of the boundaries separating neighboring chromosomal territories \cite{cremer2001,cremer2010}, determining whether they are sharp, with little mutual penetration, or are sufficiently intermixed. This question, which lies at the interface of polymer topology and chromosome architecture, deserves further investigation.

\section{Definitions, Models and Simulation Algorithms}
\label{sect:02}

To formalize the statistics of unentangled directed polymer braids ("fibers"), we map the geometric trajectories of non-crossing polymer chains onto stochastic processes defined on groups. In this section, we introduce the braid group, define random braids and topological Brownian bridges, discuss the mapping onto algebraic word reduction problems, and consider the locally-free group $LF_n$ as an analytically tractable approximant of the braid group $B_{n+1}$.

\subsection{Braids, random braids, and topological Brownian bridges (tBb)}

\begin{figure}[ht]
    \centering
    \includegraphics[width=0.7\linewidth]{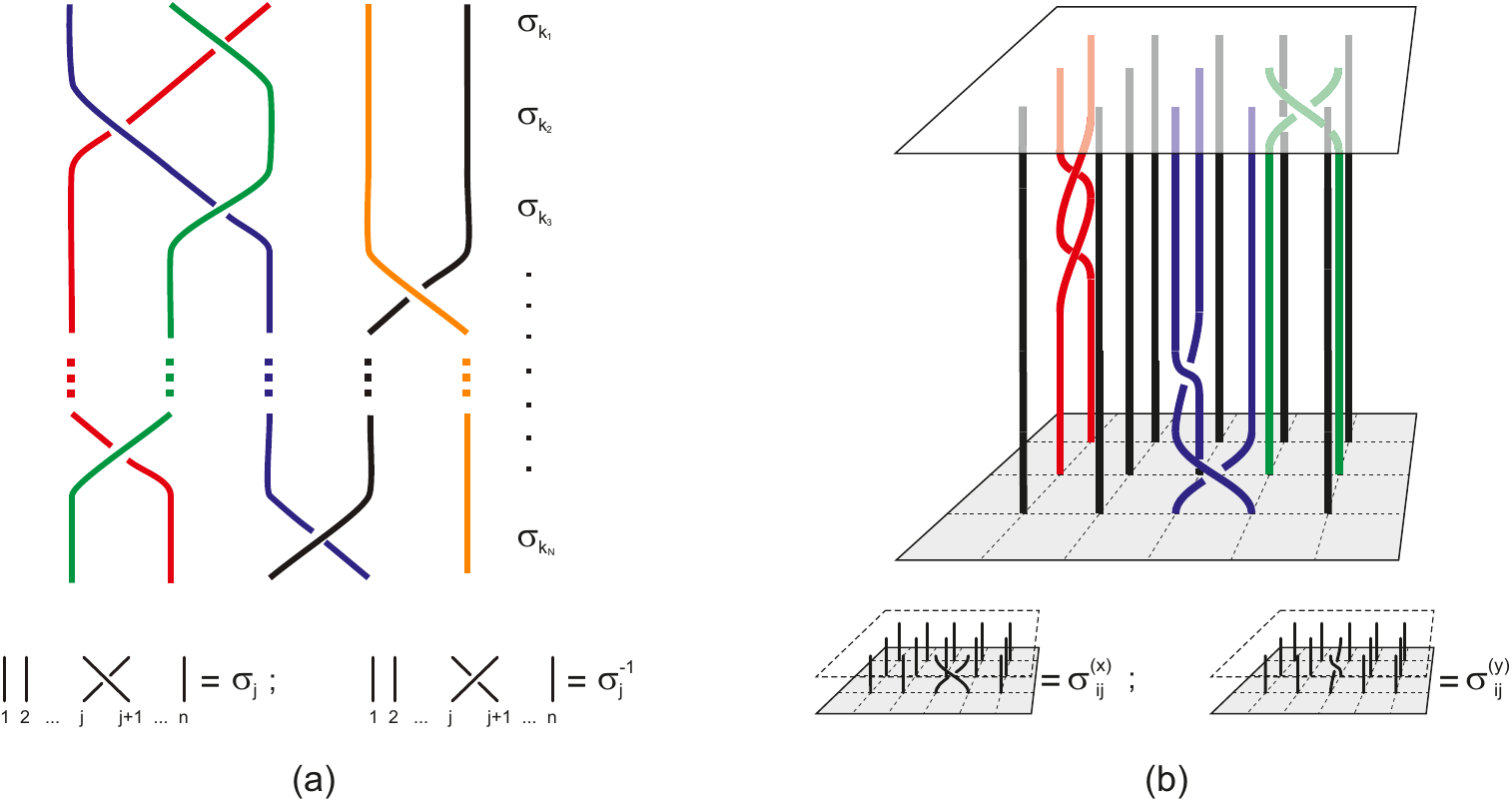}
    \caption{Schematic representation of (a) two-dimensional Artin braid generators $\sigma_i$ and (b) three-dimensional surface braid operators.}
    \label{fig:braids01}
\end{figure}

Consider a collection of $n+1$ directed, woven non-phantom strands running along a longitudinal axis $y \in [0, N]$ inside a 2D or 3D slabs between two horizontal parallel boundary lines (planes) as shown in \fig{fig:braids01}a,b for models in 2D and in 3D correspondingly. Since each strand cannot pass through any other, every configuration of trajectories with fixed endpoints splits into a distinct topological equivalence class classified by the Artin braid group $B_{n+1}$ \cite{Artin1947}.

\begin{figure}[ht]
    \centering
    \includegraphics[width=0.7\linewidth]{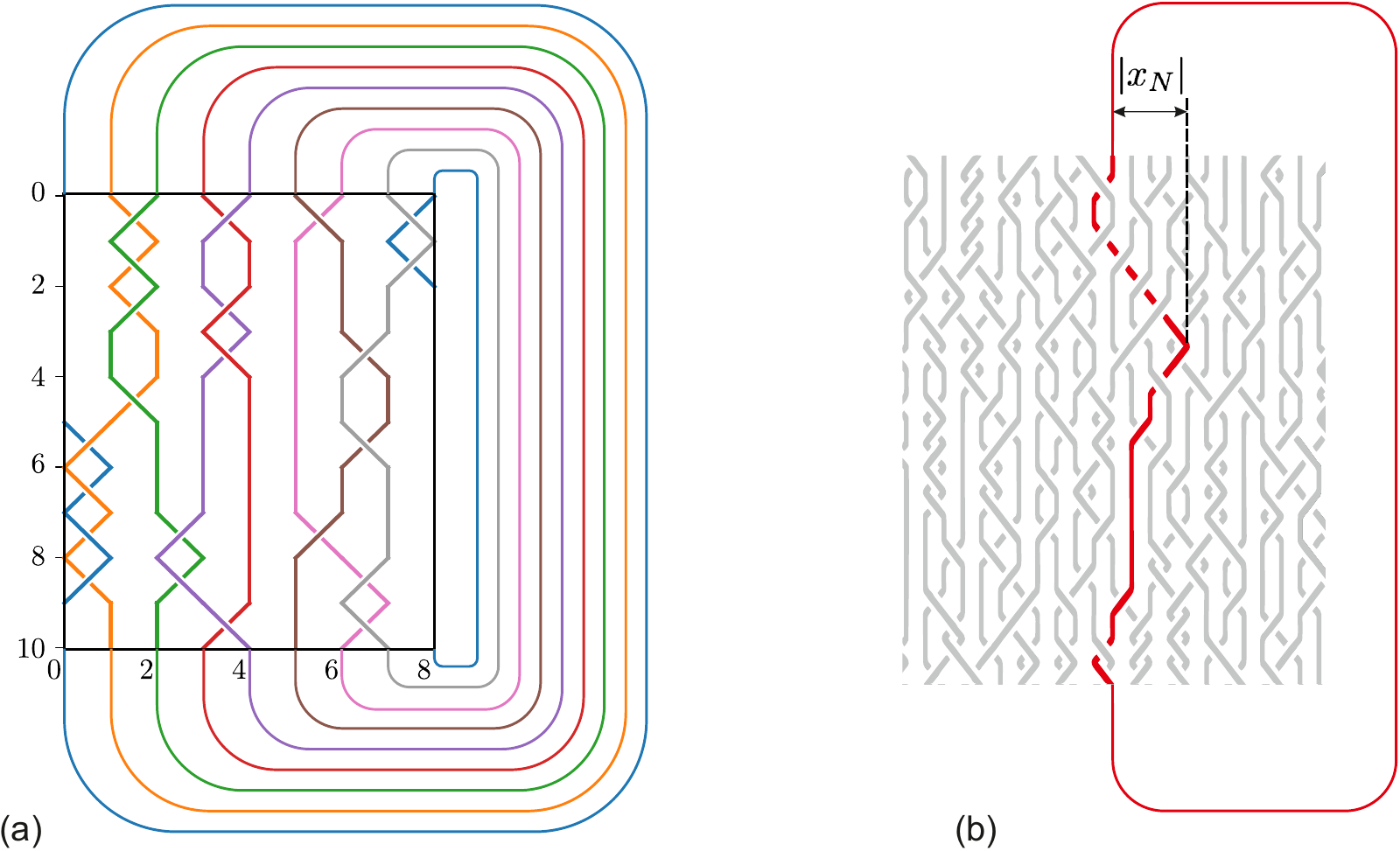}
    \caption{(a) Cylindrical closure of a $n$-strand braid. (b) Definition of the transverse span $|x_N|$ for a topological Brownian bridge.}
    \label{fig:braids02}
\end{figure}

Geometrically the topology of a braid is defined on the longitudinal interval $y \in [0,N]$, together with a closure condition that connects the open ends of each thread on a cylinder. Specifically, the terminal endpoints at $y=N$ are connected to the corresponding initial endpoints at $y=0$ by non-intersecting closure arcs, as illustrated in \fig{fig:braids02}a. Under this closure, the trivial braid sector consists of configurations whose corresponding braid group element reduces to the identity, $e$. Each thread then closes into an unknotted and unentangled ring, as shown in \fig{fig:braids02}b. Thus, the trivial sector comprises precisely a collection of $n+1$ mutually unentangled, individually unknotted closed loops.

Algebraically the braid group $B_{n+1}$ is defined by $n$ elementary generators $\{\sigma_1, \sigma_2, \dots, \sigma_n\}$, where $\sigma_j$ represents an "over-crossing" of the $i$-th and $(i+1)$-th adjacent strands along the longitudinal axis $y$, while $\sigma_j^{-1}$ denotes an "under-crossing". In 2D these generators are visualized in \fig{fig:braids01}a and obey the Artin braid relations \cite{Artin1947}:
\be
\left\{
\begin{array}{rcll}
\sigma_i \sigma_j & = & \sigma_j \sigma_i & \text{for } |i - j| \ge 2, \medskip \\
\sigma_i \sigma_{i+1} \sigma_i & = & \sigma_{i+1} \sigma_i \sigma_{i+1} & \text{for } 1 \le i \le n-1 \medskip \\
\sigma_i \sigma_i^{-1} & = & \sigma_i^{-1} \sigma_i = e & \text{for } 1 \le i \le n
\end{array}
\right.
\label{eq:artin}
\ee
where $e$ is the group unity. The first equation in \eq{eq:artin} expresses the far-commutativity of spatially separated strands, whereas the second encodes the three-strand topological interaction associated with the $3^{\rm rd}$ Reidemeister move \cite{reidemeister}. These relations can be generalized to the three-dimensional case of ``surface braid groups'' \cite{birman, paris}, whose generators $\sigma_{ij}^x$ and $\sigma_{ij}^y$ are schematically depicted in \fig{fig:braids01}b. Since the present paper is concerned exclusively with two-dimensional braids, we do not discuss surface braid groups further and leave their consideration for future work.

A random braid of length $N$ represents a discrete trajectory generated by a random walk on $B_{n+1}$. For non-biased symmetric random walk at each longitudinal step $y \to y+1$, a generator $\sigma_i^{\pm 1}$ is selected independently and uniformly from the alphabet $A=\{\sigma_1^{\pm 1}, \dots, \sigma_n^{\pm 1}\}$, yielding a formal algebraic word:
\be
w_N = \sigma_{i_1}^{\epsilon_1} \sigma_{i_2}^{\epsilon_2} \dots \sigma_{i_N}^{\epsilon_N}, \quad \epsilon_k = \pm 1.
\label{eq:word}
\ee
The canonical algebraic description of a braid word \eq{eq:word} is sequential since it has a natural "sparsity" in the longitudinal dynamics. In the "sequential updating" framework, each discrete step $y \to y+1$ accommodates only a single elementary strand crossing $\sigma_i^{\pm 1}$ across the entire bundle of $n+1$ threads, leaving all remaining strands idle at that $y$--slice.

To describe a physically "dense" polymer melt where thermal fluctuations occur simultaneously throughout the fiber bundle, one exploits the fundamental far-commutativity relation $\sigma_i \sigma_j = \sigma_j \sigma_i$ for $|i-j| \ge 2$. Because operators acting on spatially disjoint pairs of strands commute, independent strand crossings occurring at adjacent sequential steps can be freely slide along the $y$--axis and grouped into the same spatial slice without altering the topological state of the braid. This far-commutativity allows one to reorganize the sequential word $w_N$ into a sequence of maximal commuting blocks, converting the sequential updating into a parallel one. Up to the definition of the underlying group, this construction is directly analogous to the heap representation of Viennot \cite{viennot1986}  and Cartier-Foata trace monoids \cite{cartier1969}. In this "parallel update representation", up to $\lfloor (n+1)/2 \rfloor$ independent crossings can occur simultaneously within a single longitudinal layer $y$, transforming the sparse sequential walk into a model of a fluctuating unentangled directed polymers in a melt. While the sequential framework is more convenient for the analytical treatment of random walks on groups, the parallel representation is particularly useful for implementing geometric transformations in numerical algorithms.

A Brownian bridge on $B_{n+1}$, also called a "topological Brownian bridge" (tBb) is a random braid of length $N$ conditioned to return to the group identity element $e \in B_{n+1}$ at the last step $y=N$: $w_N \sim e \quad \text{in } B_{n+1}$. Geometrically, a tBb represents the trivial braid sector which corresponds to $n+1$ threads anchored at $y=0$ and $y=N$ that can be continuously transformed by Reidemeister moves into a set of parallel straight lines. The ensemble of such trivial threads defines the fully unentangled directed polymer fiber.

Every spatial "physical" trajectory of $n+1$ random directed polymers is represented by a random word $w_N$ of length $N$ in the alphabet $A$ defined above. Local thermal fluctuations of "physical threads" induce over- and under-crossings that algebraically correspond to local word transformations: either (i) a pair cancellations, $\{\sigma_i \sigma_i^{-1}~{\rm or }~\sigma^{-1}_i \sigma_i\} \to e$, or (ii) a pair creation, $e\to \{\sigma_i \sigma_i^{-1}~{\rm or }~\sigma_i^{-1} \sigma_i\}$. The constraint that the bundle of threads remains globally unentangled requires that the word $w_N$ reduces identically to $e$ under the group relations \eq{eq:artin}.

The partition function $Z(N, n)$ of the unentangled polymer melt can be formally written as:
\be
Z(N, n) = \sum_{w \in \{A^N\}} \delta\left(w_N, e\right)
\label{eq:partition}
\ee
where $\delta\left(w_N, e\right) = 1$ if the word $w_N$ reduces to the identity, $e$, and $\delta\left(w_N, e\right) = 0$ otherwise; the summation runs over the alphabet $A$ on each on $N$ steps ("slices"), $y\to y+1$, where $y=0,1,...,N-1$. The transverse span (i.e. excursion along the $x$-direction) of an individual thread in a braid of length $N$ is $|x_N|$ in \fig{fig:braids02}b. It can be estimated by the typical shortest length of a half-braid within the topological Brownian bridge ensemble. We explicitly construct this mapping in the following section considering braid in the "locally-free" approximation \cite{vershik2000}.

\subsection{The locally-free group $LF_n$ as a bilateral approximation of a braid group $B_{n+1}$}

Solving the word problem in the full braid group $B_{n+1}$ for long polymers ($N \gg 1$) and large bundles ($n \gg 1$) is computationally challenging due to the non-triviality of Garside normal forms and the high complexity of braid reduction algorithms \cite{garside1969, dehornoy2000}. To bypass this difficulty, we model $B_{n+1}$ using the locally-free group $LF_n$ \cite{vershik2000}.

The locally-free group $LF_n$ is defined by $n$ generators $\{g_1, g_2, \dots, g_n\}$ subject strictly to far-commutativity relations:
\be
\left\{
\begin{array}{rcl}
g_i g_j & = & g_j g_i \quad \text{for } |i - j| \ge 2, \medskip \\
g_i g_i^{-1} & = & g_i^{-1}g_i = e
\label{eq:loc-fr}
\end{array}
\right.
\ee
with no relations imposed on adjacent generators $g_i$ and $g_{i+1}$ ($i = 1, \dots, n-1$). Consequently, any pair of adjacent generators generates a free subgroup $F_2 \subset LF_n$, which gives the group its name \cite{vershik2000}. 

To avoid potential confusion regarding indexing, we emphasize that in the standard notation for the braid group $B_{n+1}$, the subscript specifies the total number of strands $(n+1)$, which corresponds to $n$ elementary generators $\{\sigma_1, \dots, \sigma_n\}$. In contrast, for the locally-free group $LF_n$, the index $n$ directly counts the number of generators $\{g_1, \dots, g_n\}$.

The group $LF_n$ provides a bilateral approximation (a "sandwiching") for word enumeration in $B_{n+1}$ (see \cite{vershik2000} for details): 
\begin{itemize}
\item On one hand, $LF_n$ has fewer relations than $B_{n+1}$ because it omits the non-abelian triple braid-like relation. As a result, word simplification in $LF_n$ is governed by local inverse cancellations ($g_i g_i^{-1} \to e$) combined with commutation of distant generators. This makes $LF_n$ algebraically larger than $B_{n+1}$.
\item On the other hand, it is known that the generators $g_i$ of $LF_n$ and $\sigma_i$ of $B_{n+1}$ obey the relation $g_i = \sigma_i^2$, which identifies each $g_i$ with a full monodromny of adjacent strands $i$ and $i+1$. The generator $g_i$ acts on the neighboring strands retaining their identity permutations (as in pure braids), while capturing essential local topological entanglements. Hence, $LF_n$ is embedded into $B_{n+1}$.
\end{itemize}

Thus, number of nonequivalent words representing different topological sectors of the braid of irreducible length $2k$ in the braid group $B_{n+1}$, which we call "the volume", $V_{B_{n+1}}(2k)$, is sandwiched between the number of irreducible words of length $k$ and the number of irreducible words of length $2k$ in the locally free group $LF_n$:
\be
V_{LF_n}(k)\le V_{B_{n+1}}(2k)\le V_{LF_n}(2k)
\label{eq:volume}
\ee
By mapping $LF_n$ onto Viennot's theory of heaps of pieces \cite{viennot1986}, the word enumeration problem can be solved via the spectral analysis of the corresponding transfer matrix. 

The growth of the locally free group volume, $V_{LF_n}(k)$, with reduced word length, $k$, reveals the asymptotic geometric structure of the Cayley graph of the group $LF_n$. Analyzing the statistics of topological Brownian bridges on the Cayley graph allows us to derive theoretical estimates for the transverse polymer span $|x_N|$, thereby establishing a foundation for topological localization. 

Below we investigate, both numerically and analytically, the transverse fluctuations of individual strands in a melt of unentangled and unknotted directed polymers within the locally-free approximation. While specific numerical prefactors are model-dependent, the generic scaling behavior and our conclusions regarding topological localization are independent of approximating the braid group by the locally-free group.

\subsection{Models and simulation algorithms}

We consider four models of directed polymer fibers: (i) the single unconstrained directed polymer (a null model), (ii) the single directed polymer in a regular array of parallel straight lines, (iii) the single directed polymer in a gaped array of parallel straight lines, and (iv) the fully interacting melt of unentangled directed polymers in a locally-free approximation. The models (i)--(iv) differ in their geometries and in the constraints imposed on the accessible configurations, but all share the following common features:

\begin{itemize}
\item Each fiber has a fixed longitudinal length $N$ along $y$--coordinate;
\item Transverse motion is represented by the generators $\sigma_i(y)$ and $\sigma_i^{-1}(y)$. The identity operator denotes the absence of transverse motion between layers $y$ and $y+1$.
\item The system is periodic in the longitudinal $y$--direction, $x_i(0)=x_i(N)$ for each $i\in [1,n]$ (i.e. upper and lower open ends of strings are glued). Each fiber therefore forms a $y$--periodic random walk in the $(x,y)$ plane, subject to model-dependent constraints described below. The configuration of each of $n$ stands is a topologically trivial Brownian bridge (unknotted and unentangled ring). Algebraically, it means that the operator word associated with each configuration by repeated application of the relevant local relations is reducible to the identity operator.
\item Each simulation is initialized in a maximally extended configuration,
\be
\begin{cases}
x_i(0)-x_i(N/2) = N/2 \mbox{ for even $i$}, \medskip \\
x_i(0)-x_i(N/2) = -N/2 \mbox{ for odd $i$}.
\end{cases}
\ee
For models (i)--(iii) we regard the ensemble as equilibrated, once the mean transverse displacement no longer retains a directional bias inherited from the initial condition for, while for the model (iv) the equilibration condition is slightly different (see below).
\end{itemize}

The models (i)--(iv) are introduced in order of increasing complexity from a simple directed Brownian bridge, to a fully interacting braid of directed random walks. The results of simulations of all these models are collected in Section \ref{sect:03}.

\subsubsection{Model 1: Single directed polymer in a free space}
\label{sect:02-1}

Here, we describe a null model of a simple directed polymer forming a Brownian bridge on a line, treating it as a special case of a broader model: a directed polymer in a gaped array of uncrossable lines where the gap size tends to infinity. This framework provides a unified description that we apply throughout our analysis of all models. 

For a $y$--periodic single directed polymer, the local operators are $\sigma_i(y)$, $\sigma_i^{-1}(y)$, $e(y)$, where $i\in\{1,2\}$. The operator $\sigma_i(y)$ produces a unit step in the positive $x$-direction between layers $y$ and $y+1$, whereas $\sigma_i^{-1}(y)$ produces a unit step in the negative $x$-direction. The indices $i=1$ and $i=2$ have no distinct geometric meaning in this null model because of absence of any topological constraints, but are retained to facilitate comparison with the models considered in next two sections. 

\begin{figure}[ht]
    \centering
    \includegraphics[width=\linewidth]{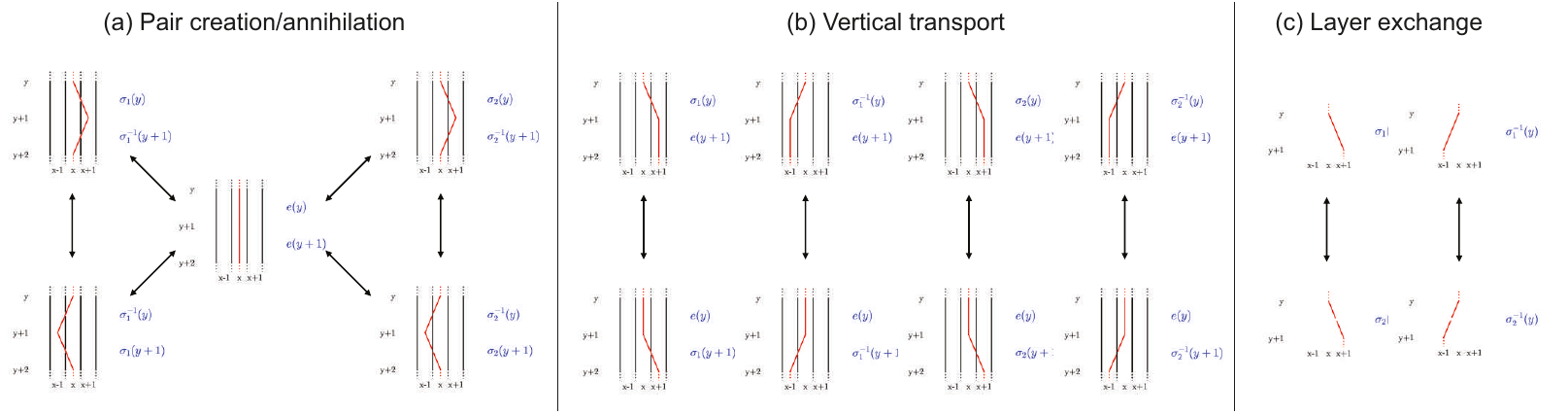}
    \caption{Monte Carlo moves for the simple directed Brownian bridge. The red line denotes the mobile polymer fiber, the black lines denote stationary "topological obstacles", and the corresponding generators are shown in blue. (a) Pair creation and annihilation inserts or removes a generator and its inverse while preserving topological triviality between layers $y$ and $y+2$. (b) Vertical transport moves a crossing by one layer through its commutation with an identity element. (c) Layer exchange is allowed only in obstacle-free regions, where changing an over-crossing into an under-crossing, or vice versa, does not introduce an additional topological constraint. Double-headed arrows denote reversible moves.}
    \label{fig:noninteracting-braid-moves}
\end{figure}

A directed polymer configuration is represented by an operator word of length $N$ and is sampled using the generic set of moves shown in \fig{fig:noninteracting-braid-moves}. Pair creation and annihilation preserve longitudinal periodicity, vertical transport moves generators along the $y$-direction, and layer exchange permits transitions between the two generator labels. The move set is intentionally minimal to connect to the systems described below and may be augmented to improve mixing efficiency.

The maximally extended initial state is constructed by alternating $\sigma_1(y)$ and $\sigma_2(y)$ for $y<N/2$ and filling the remaining layers with the corresponding inverse generators. For example, for $N=10$ we start with the configuration $[\sigma_1,\sigma_2,\sigma_1,\sigma_2,\sigma_1, \sigma_1^{-1},\sigma_2^{-1},\sigma_1^{-1},\sigma_2^{-1},\sigma_1^{-1}]$.

\subsubsection{Model 2: Single directed polymer in a regular array of straight lines}
\label{sect:02-2}

\begin{figure}[ht]
    \centering
    \includegraphics[width=0.65\linewidth]{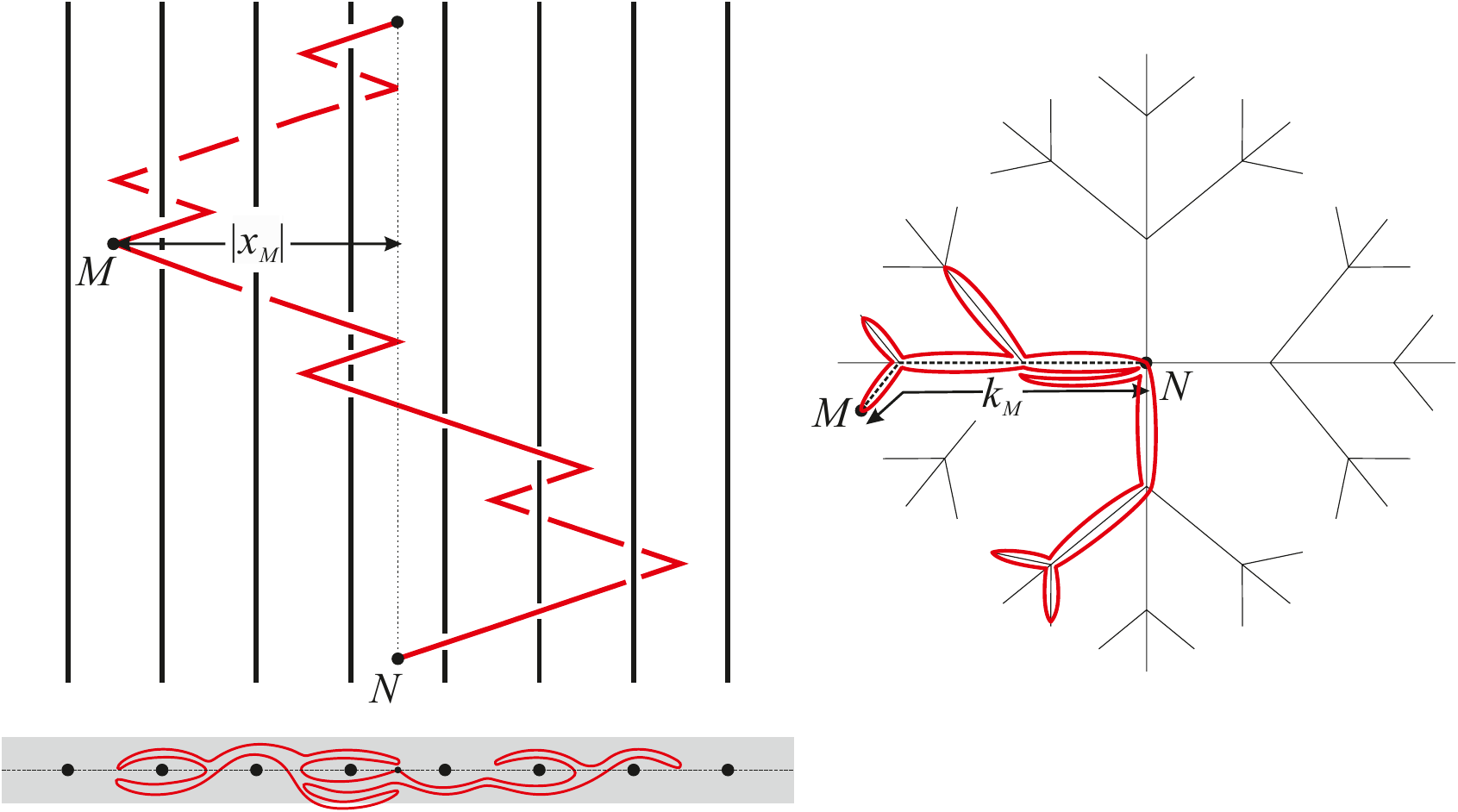}
    \caption{Topological Brownian bridge in a regular 1D lattice of obstacles (left) and its mapping onto a random walk on a Cayley tree (right).}
    \label{fig:braids04}
\end{figure}

The second model describes a topological Brownian bridge -- the single directed polymer fiber fluctuating in an array of immobile straight lines. The model itself and its schematic top view are shown in the left panel of \fig{fig:braids04}. Whenever the polymer changes its transverse coordinate while advancing in the $y$-direction, it must pass either over or under an obstacle whose $x$-position is fixed and which extends through the full longitudinal length of the system. This model represents the 1D version of the model "polymer chain in the lattice of obstacles studied in \cite{KhokhlovNechaev1985, helfand-pearson, rubinstein-helfand}. It was shown in \cite{KhokhlovNechaev1985} that a random walk in a lattice of obstacles corresponds bijectively to an unconstrained random walk on the Cayley tree (shown in the right panel of \fig{fig:braids04}) or, in the continuous case, to a random walk on a modular form that can be explicitly constructed in terms of Jacobi elliptic functions \cite{nechaev1988, nechaev1999}.

The local operators are again $\sigma_i(y)$, $\sigma_i^{-1}(y)$, $e(y)$, with $i\in\{1,2\}$. Here, however, the labels of generators have distinct geometric meanings: $\sigma_1$ denotes a crossing over an obstacle and $\sigma_2$ denotes a crossing under an obstacle. The sign of the exponent retains the directional convention introduced in Section \ref{sect:02-1}.

Topological triviality imposes the local relations
\be
\left\{\begin{array}{rcl}
\sigma_i(y)\sigma_j^{-1}(y+1) & = & \sigma_j^{-1}(y)\sigma_i(y+1) = e(y)e(y+1) \mbox{ for $i=j$}, 
\medskip \\ \sigma_i(y)e(y+1) & = & e(y)\sigma_i(y+1), \medskip \\
\sigma_i^{-1}(y)e(y+1) & = & e(y)\sigma_i^{-1}(y+1) \end{array}\right.
\ee
The corresponding Monte Carlo moves are shown in \fig{fig:noninteracting-braid-moves}a,b. The initial condition is the same as in Section \ref{sect:02-1}.

\subsubsection{Model 3: Single directed polymer in a gaped array of straight lines}
\label{sect:02-3}

\begin{figure}[ht]
    \centering
    \includegraphics[width=0.5\linewidth]{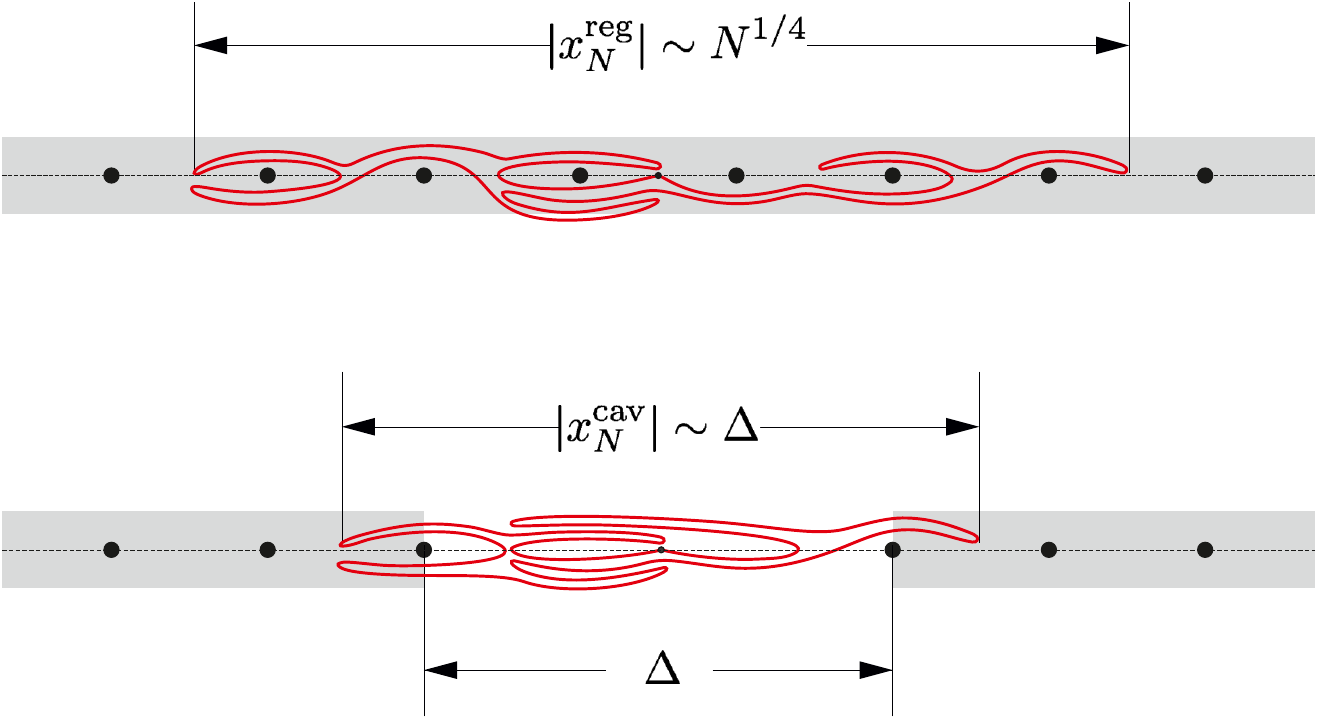}
    \caption{A directed polymer thread fluctuating in a gaped array of topological obstacles featuring a defect cavity of width $\Delta$.}
    \label{fig:braids05}
\end{figure}

The Model 3 described earlier in \cite{nechaev-rev}, interpolates between the unconstrained Brownian bridge and the topologically trivial Brownian bridge in a regular array of immobile lines. We introduce a line-free cavity ("hole") of width $\Delta$ within a uniform forest of vertical lines -- see \fig{fig:braids05}. Inside this cavity, centered around $x=0$, the over- and under-passages are indistinguishable and are subject to exchange, as in Model 1.

We therefore use all moves shown in \fig{fig:noninteracting-braid-moves}, but permit the layer-exchange move shown in \fig{fig:noninteracting-braid-moves}c only when $|x(y)|\leq \Delta/2$. The initial conditions remain unchanged.

\subsubsection{Model 4: Fully interacting melt of unentangled directed polymers in a locally-free approximation}
\label{sect:02-4}

\begin{figure}[ht]
\centering
\includegraphics[width=0.75\linewidth]{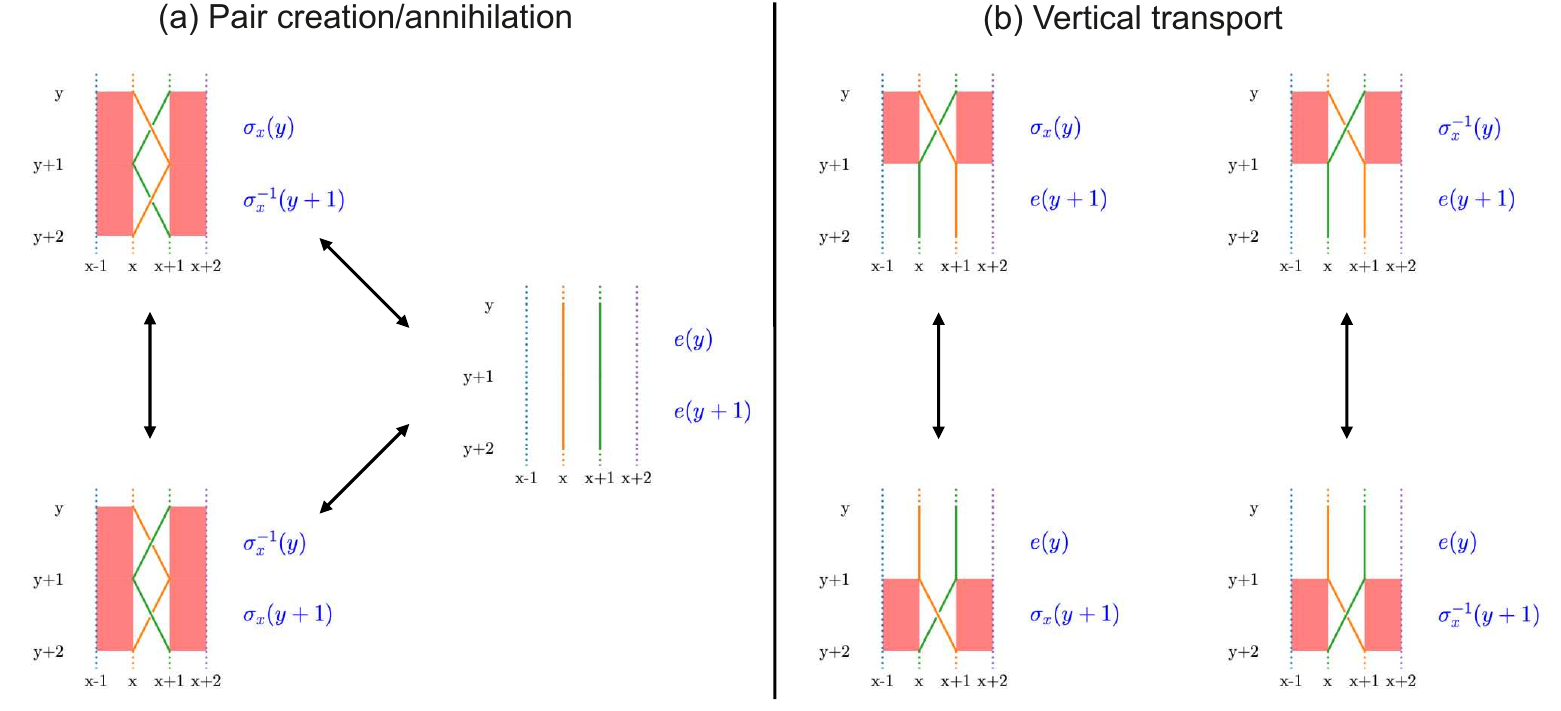}
\caption{Monte Carlo moves used for the Model 4. Colored lines represent different polymer fibers. Red-shaded cells indicate locations that are inaccessible because two crossings cannot occupy adjacent positions. The corresponding local operators are shown in blue. (a) Pair creation and annihilation insert or remove a generator and its inverse while preserving topological triviality between layers $y$ and $y+2$. (b) Vertical transport moves a crossing by one layer through its commutation with an identity element. Double-headed arrows denote reversible moves.}
\label{fig:interacting-braid-moves}
\end{figure}

To generalize the preceding models to a fully interacting braid, we introduce the local operators $\sigma_x(y)$, $\sigma_x^{-1}(y)$, $e_x(y)$, where the subscript $x$ identifies the transverse bond on which the operator acts. Periodic boundary conditions are imposed in the transverse direction, so the system may be viewed as a collection of rings on a torus. Algebraically, the braid is represented by a two-dimensional operator-valued lattice in which each cell at $(x,y)$ contains one local operator -- see \fig{fig:interacting-braid-moves}. We vary both the longitudinal length $N$ and the transverse system size defined by the number of fibers, $n$.

The polymer configuration is reconstructed from the operator lattice as follows:
\begin{itemize}
\item $\sigma_x(y)$ denotes a crossing in which the fiber at $(x,y)$ advances to $(x+1,y+1)$ over the fiber at $(x+1,y)$, which advances to $(x,y+1)$,
\item $\sigma_x^{-1}(y)$ and $\sigma_x^{-1}(y)$ denote correspondingly over-- and under--crossing,
\item $e_x(y)$ denotes the absence of a crossing between two fibers as they advance from layer $y$ to layer $y+1$.
\end{itemize}

The local relations are
\be
\left\{\begin{array}{rcll}
\sigma_x(y)\sigma_{x'}^{-1}(y+1) & = & \sigma_{x'}^{-1}(y)\sigma_x(y+1) = e_x(y)e_{x'}(y+1) & \mbox{ for $x=x'$} \medskip \\
\sigma_x(y)e_{x'}(y+1) & = & e_{x'}(y)\sigma_x(y+1) & \mbox{ for $x=x'$} \medskip \\
\sigma_x^{-1}(y)e_{x'}(y+1) & = & e_{x'}(y)\sigma_x^{-1}(y+1) & \mbox{ for $x=x'$} \medskip \\
\sigma_x(y)\sigma_{x'}(y+1) & = & \sigma_{x'}(y)\sigma_x(y+1) & \mbox{ for $|x'-x|\geq 2$} \medskip \\
\sigma_x^{-1}(y)\sigma_{x'}^{-1}(y+1) & = & \sigma_{x'}^{-1}(y)\sigma_x^{-1}(y+1) & \mbox{ for $|x'-x|\geq 2$} \medskip \\ 
\sigma_x(y)\sigma_{x'}^{-1}(y+1) & = & \sigma_{x'}^{-1}(y)\sigma_x(y+1) & \mbox{ for $|x'-x|\geq 2$}
\end{array} \right.
\label{eq:commutator4}
\ee

These relations allow generators acting on well-separated transverse bonds to commute, while crossings involving neighboring fibers remain constrained. To speed up numerical algorithm we do not impose the full braid-group relations that would permit a crossing to move continuously past an adjacent crossing. Instead, we consider the braid in the locally-free approximation which means elimination of the triple ("Yang-Baxter") braid relation $\sigma_i\sigma_{i+1}\sigma_i = \sigma_{i+1}\sigma_i\sigma_{i+1}$. The relationship between the locally free group used here and the conventional braid group was discussed in Section \ref{sect:02}, while computational details are presented in Section \ref{sect:04}.

The Monte Carlo moves are shown in \fig{fig:interacting-braid-moves}. Together with the geometric interpretation of the operators, the local relations prohibit crossings from occupying adjacent transverse positions. The corresponding excluded cells are indicated by the red shaded regions in \fig{fig:interacting-braid-moves}. 

\begin{figure}[ht]
    \centering
    \includegraphics[width=0.8\linewidth]{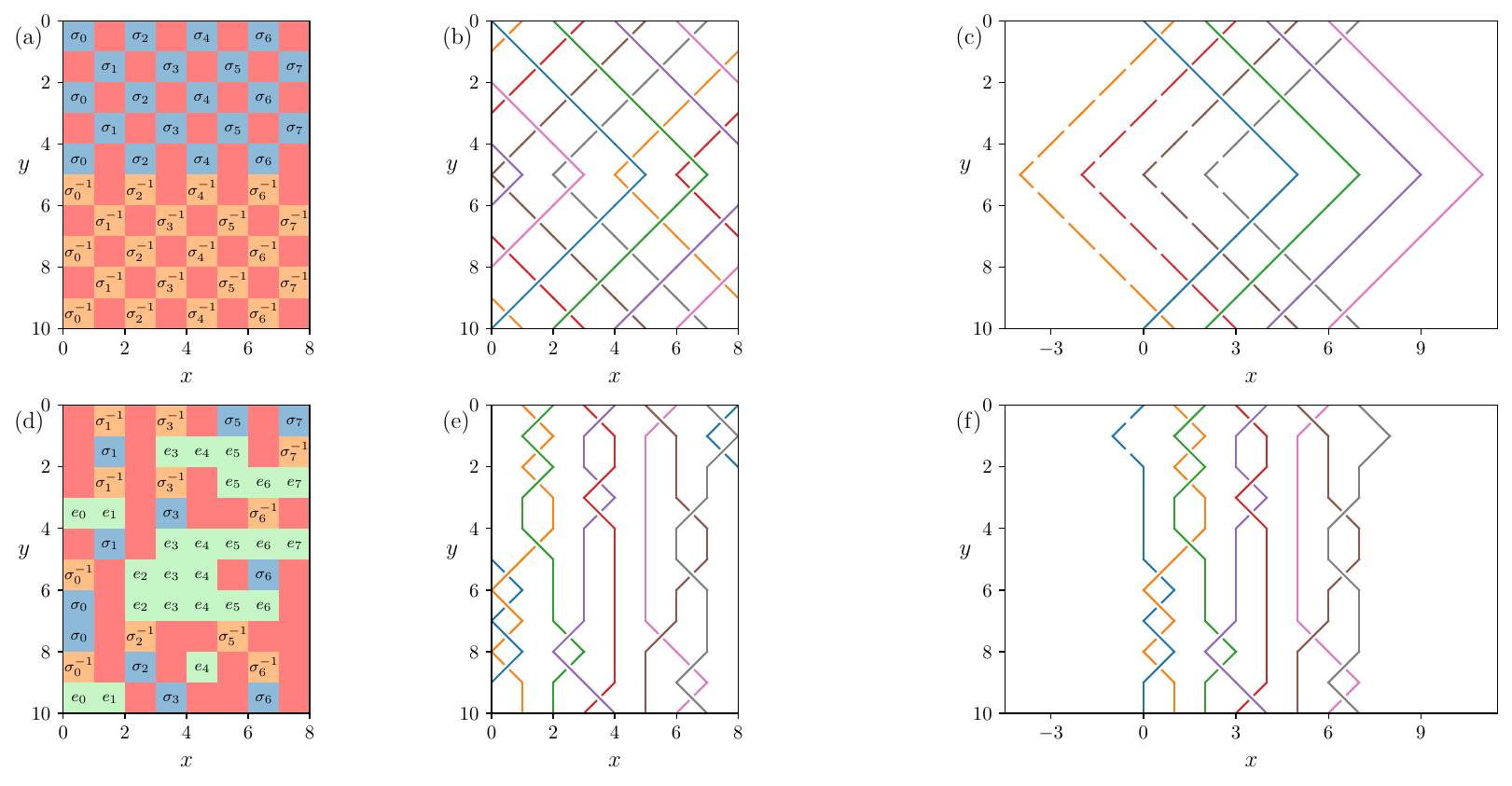}
    \caption{A fully interacting braid with length $N=10$ and $n=8$ fibers, shown in the initial state [panels (a)--(c)] and after equilibration [panels (d)--(f)]. Panels (a) and (d) show the operator lattice used in the Monte Carlo simulation. Blue cells denote $\sigma_x(y)$, orange cells denote $\sigma_x^{-1}(y)$, light-green cells denote $e_x(y)$, and red cells denote locations excluded by \eq{eq:commutator4}. Panels (b) and (e) show the corresponding fibers in periodic coordinates, while panels (c) and (f) show the same configurations in unwrapped coordinates.}
    \label{fig:braid_initial_final}
\end{figure}

The initial state is constructed from an alternating pattern of $\sigma_x$ operators that repeats diagonally across the upper half of the operator lattice -- see \fig{fig:braid_initial_final}. The lower half is filled with the corresponding inverse operators. This construction guarantees topological triviality while maximizing the initial transverse extension of every fiber.

\section{Simulation results}
\label{sect:03}

We have simulated using Monte-Carlo algorithm fibers with longitudinal lengths $N=\{10,100,300,600,1200\}$. The transverse wandering is characterized by the absolute value of the displacement of the middle point, $y=N/2$, of a threads, i.e. $\la x(N/2)|\ra$, averaged over 4800 independently generated samples. Recall that all threads are periodic in both the $x$-- and $y$--directions, and the initial conformation is chosen to be as far from equilibrium as possible. Upon release, the fibers begin to equilibrate. Equilibration is considered complete when the drift toward equilibrium vanishes (i.e., begins to fluctuate randomly in sign). This criterion is applied to Models 1--3; however, for Model 4, the relaxation is exponentially slow, so the simulation is terminated when the relative drift falls below $10^{-4}$. 

\begin{figure}[ht]
    \centering
    \includegraphics[width=0.6\linewidth]{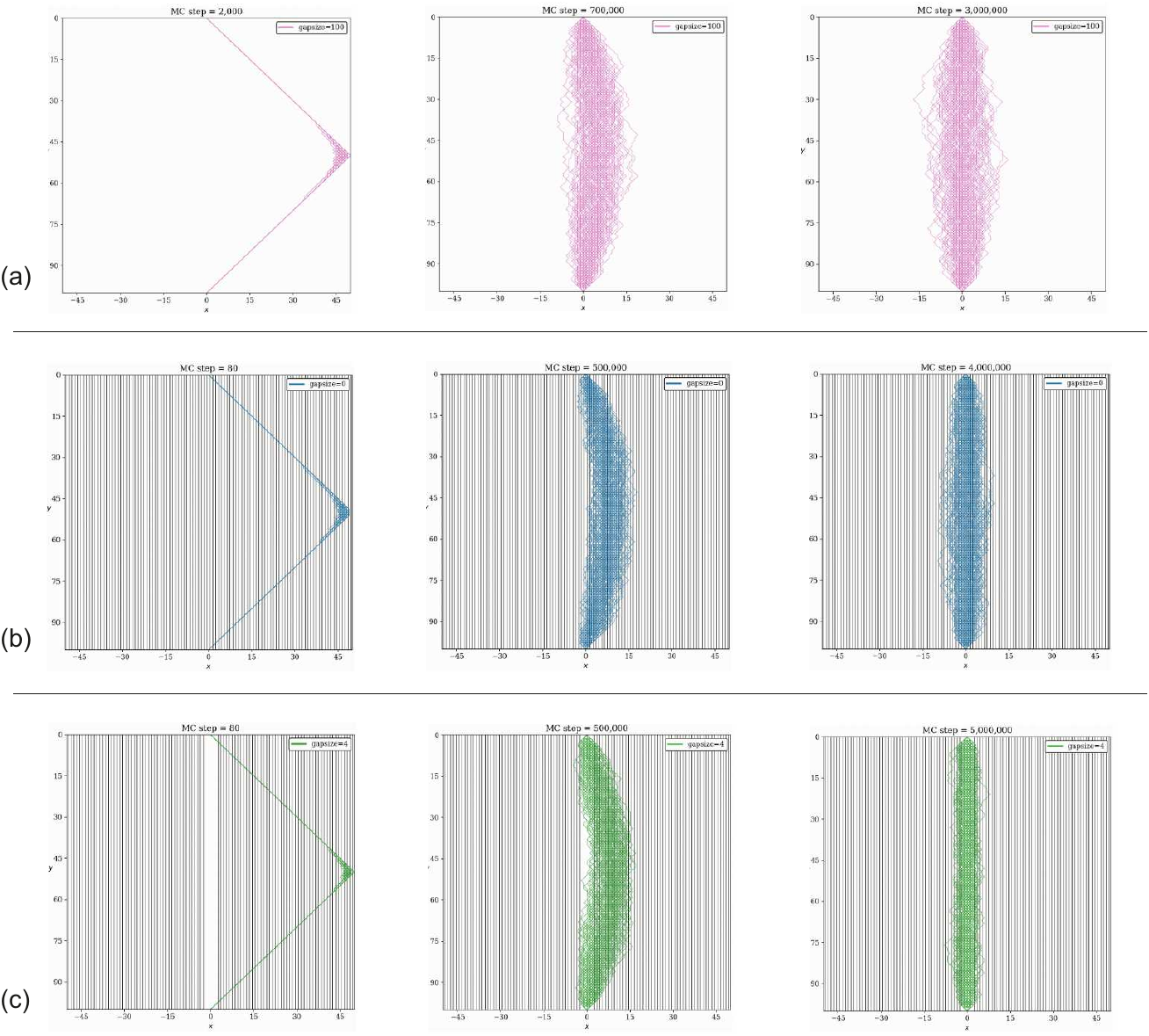}
    \caption{Sequential snapshots of relaxation of a thread from maximally expanded unentangled configuration for Models 1-3. Equilibrium behavior: (a) $\la|x_N|\ra\sim N^{1/2}$ for Model 1; (b) $\la|x_N|\ra\sim N^{1/4}$ for Model 2; (c) $\la|x_N|\ra\sim \Delta$ for Model 3, where $\Delta$ is the size of a gap.}
    \label{fig:moves-array}
\end{figure}

The snapshots of different stages of equilibration for models 1-3 are shown in \fig{fig:moves-array}a-c. Note shrinking of a Brownian bridge in a small gap of size $\Delta=4$ in last panel of \fig{fig:moves-array}c. The corresponding animations are provided in separate files \textrm{\bf fiber-empty.gif, fiber-gap0.gif, fiber-gap4.gif}.

\begin{figure}[ht]
    \centering
    \includegraphics[width=0.8\linewidth]{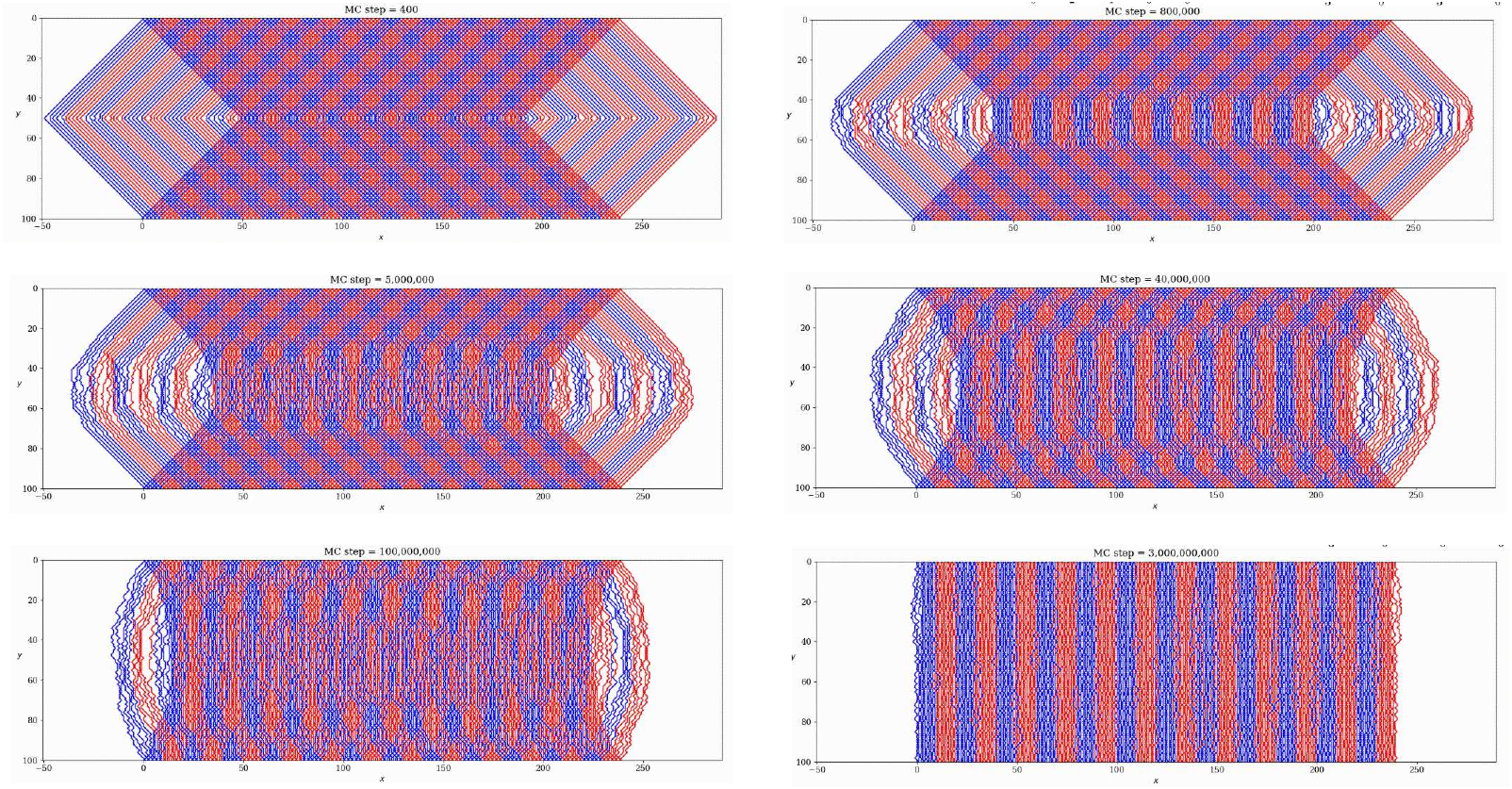}
    \caption{Sequential snapshots of the relaxation of the fiber from its maximally expanded unentangled configuration. The relaxation, which employs the minimal set of elementary moves compatible with the Reidemeister moves, proceeds through a series of instabilities where partially aligned configurations interlace with mixed ones.}
    \label{fig:moves-fiber}
\end{figure}

The snapshots of different stages of equilibration for model 4 is shown in \fig{fig:moves-fiber}. The corresponding animation is provided in a separate file \textrm{\bf fiber-full.gif}. We found that in course of equilibration from maximally extended configuration the braid undergoes a sequence of collective instabilities. Whether this is a typical feature of the collective behavior of the system, or just the consequence of the chosen minimal set of local moves, is yet an open question. 

\begin{figure}[ht]
\centering
\includegraphics[width=0.8\linewidth]{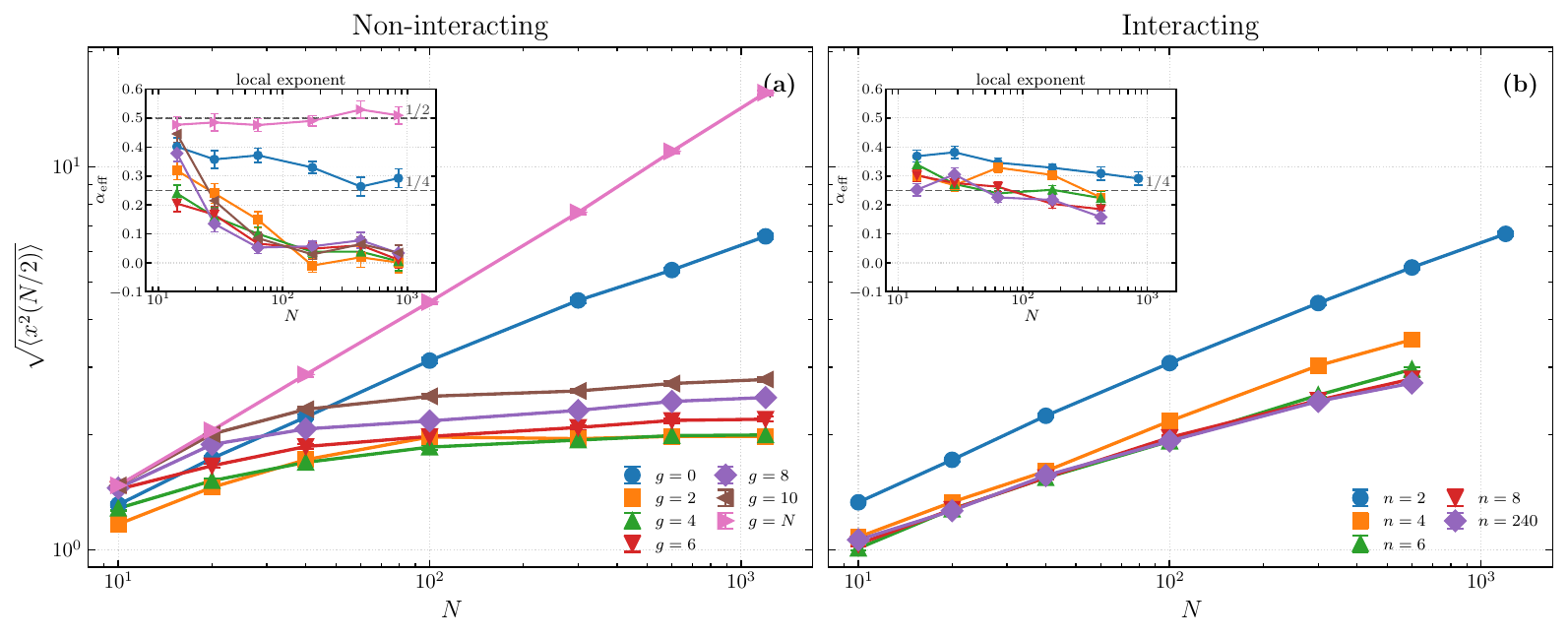}
\caption{Transverse displacement $\la|x(N/2)|\ra$ as a function of $N$. Colors denote the polymer length $N$, and marker shapes denote the model parameter. Dashed lines indicate reference slopes of $1/2$, $1/4$ and $0$. (a) Thread in a lattice of parallel lines with a central thread-free cavity of width $\Delta$. The limit $\Delta=N$ corresponds to the free Brownian bridge, whereas $\Delta=0$ corresponds to a uniform array of parallel lines. (b) Fully interacting braid in a locally-free approximation with different numbers of threads, $n$, and periodic boundary conditions along $x$- and $y$-directions.}
\label{fig:span_scaling}
\end{figure}

The simulation results corresponding to moves shown in \fig{fig:moves-array} and \fig{fig:moves-fiber} are summarized in \fig{fig:span_scaling}. Below we comment the obtained results. 

1. For the null model (Model 1), i.e. for the free Brownian bridge we get: 
\be
\la |x(N/2)|\ra \sim N^{1/2}
\ee
as expected for a periodic free random walk (simple Brownian bridge). This agreement provides a validation of the simulation procedure.

2. For a directed polymer in an uniform immobile forest of parallel lines representing the topological constraints, the transverse wandering for a single topologically trivial thread is suppressed and is consistent, over the accessible range, with the algebraic "annealed branching polymer" scaling \cite{daoud1981, KhokhlovNechaev1985}:
\be
\la |x(N/2)|\ra \sim N^{1/4}
\label{eq:regular}
\ee

3. For a directed polymer in a gaped array of straight lines a clear confinement is seen:
\be
\la |x(N/2)|\ra \sim \Delta
\ee
Larger cavities (i.e. larger $\Delta$) permit larger transverse wandering. So, the directed polymer is localized on the cavity starting from the size $\Delta =2$ measured in spacing between straight lines.

4. For the braid of two threads ($n=2$) the data are also consistent with an algebraic scaling close to $1/4$ over the shown range of lengths. As the number of threads increases, the curves bend away from this power-law and tend to confined, weakly $y$-dependent regime. Based on analytic arguments provided in next sections, we conjecture that each thread localizes in its own tube of finite width independent on $N$ in the $N\to \infty$ limit:
\be
\la |x(N/2)|\ra \sim \mathrm{const}
\ee
The plots for $n=5,6$ are already close to those for $n=240$, suggesting that the many-thread regime is reached at relatively small braid widths of order of $n\sim 5\div 6$. 

\section{Theory}
\label{sect:04}

\subsection{Flory-type approach: Localization of unentangled random walk in a gaped array of obstacles}
\label{sect:04-1}

We consider a quasi-one-dimensional random polymer loop consisting of $N \gg 1$ steps of length $a$. The loop is conditioned to be unentangled with a one-dimensional lattice of obstacles with spatial period $c > a$ and local coordination number, $z=4$ (right-over, right-under, left-over, left-under). Note that, for a regular lattice of obstacles on the line, the only possible value is $z=4$. We nevertheless consider $z$ as a free parameter to cover the case with no array of obstacles, which formally corresponds to $z=2$. The lattice contains a localized obstacle-free cavity ("defect") of size $\Delta > c$. Our goal is to demonstrate that there is a critical value $\Delta_{\rm c}$ above which the loop is localized on the cavity and does not spread in lattice of obstacles.

In the asymptotic limit $N \gg 1$, the partition functions of closed loops in the gapless array of obstacles, $W_N^{\text{arr}}$, and within the cavity, $W_N^{\text{cav}}$, are
\be
\begin{array}{l}
\disp W_N^{\text{arr}} \sim N^{-\gamma} \, 2^N (z-1)^{\frac{N a}{2c}} \exp\left(-\frac{\pi^2a^2}{2c^2}N\right) \medskip \\
\disp W_N^{\text{cav}} \sim N^{-\gamma} \, 2^N \exp\left(-\frac{\pi^2a^2}{2\Delta^2} N \right)
\end{array}
\ee
where $\gamma= 3/2$ is the universal exponent of the random walk first return in 1D. The Gaussian factor $\exp\left(-\frac{\pi^2 a^2}{2L^2} N\right)$ for $L = \{c, \Delta\}$ represents the ground-state eigenvalue of the diffusion operator $-\frac{a^2}{2}\frac{d^2}{dx^2}$ with Dirichlet boundary conditions inside a domain of width $L$. The corresponding free energies $F_{\text{arr, cav}} = -k_B T \ln W_N^{\text{arr, cav}}$ for $N$--step ($N \gg 1$) random walks are:  
\be
\begin{array}{l}
\disp F_{\text{arr}} = -k_B T N \left[ \ln 2 + \frac{a}{2c}\ln(z-1) - \frac{\pi^2}{2} \frac{a^2}{c^2} \right] + \frac{3}{2} k_B T \ln N \medskip \\
\disp F_{\text{cav}} = -k_B T N \left[ \ln 2 - \frac{\pi^2}{2} \frac{a^2}{\Delta^2} \right] + \frac{3}{2} k_B T \ln N
\end{array}
\ee

The loop localizes inside the cavity when $F_{\text{cav}} \le F_{\text{arr}}$. In this inequality, both the extensive free-walk entropy $-N \ln 2$ and the sub-extensive logarithmic loop entropy corrections $\frac{3}{2} \ln N$ cancel identically at both sides of the inequality. The remaining terms set the condition for the localization threshold which is independent of the loop length $N$ in the thermodynamic limit: 
\be
\frac{\pi^2}{2} \frac{a^2}{\Delta^2} \le \frac{\pi^2}{2} \frac{a^2}{c^2} - \frac{a}{2c}\ln(z-1)
\ee
yielding the critical cavity size $\Delta_{\rm c}$ above which localization occurs:
\be
\Delta \ge \Delta_{\rm c} = \frac{c}{\sqrt{1 - \frac{c\ln(z-1)}{\pi^2 a}}}
\ee
Physically, the localization criterion $\Delta \ge \Delta_{\rm c}$ reflects the competition between spatial confinement in a cavity and topological entropy loss of unentangled trajectories in the perfect array of obstacles. When $\Delta \ge \Delta_{\rm c}$, the entropy of spatial confinement gains of the loss of topological branching entropy, driving a localization transition of the Brownian bridge inside the gap. For $z = 2$, topological constraints disappear, leading to $\Delta_{\rm c} = c$, where any cavity larger than the bulk cell size induces spatial localization.

This result schematically demonstrates the entropic nature of localization of threads in a braid. Even small spatial cavities in the regular array of obstacles are entropically favorable for randomly fluctuating threads. So, the algebraic span of the thread in the regular lattice of topological obstacles, given by \eq{eq:regular} is unstable and is subject to demonstrate localization under small perturbation of the environment.

\subsection{Beyond the mean-field approach: Localization of topological Brownian bridges on a graph of a locally-free group}
\label{sect:04-2}

\subsubsection{Asymptotic Cayley graph of a locally-free group $LF_n$}

In this section, our primary goal is to study the statistics of Brownian bridges on the Cayley graph of the locally free group $LF_n$. Recall that a closed path on the Cayley graph of $LF_n$ corresponds to a simultaneously trivial topology of all threads. This equivalence holds because the word representing the path on the Cayley tree, expressed in terms of the generators of $LF_n$, reduces fully to the identity element under the Brownian bridge condition.

\begin{figure}[ht]
    \centering
    \includegraphics[width=0.65\linewidth]{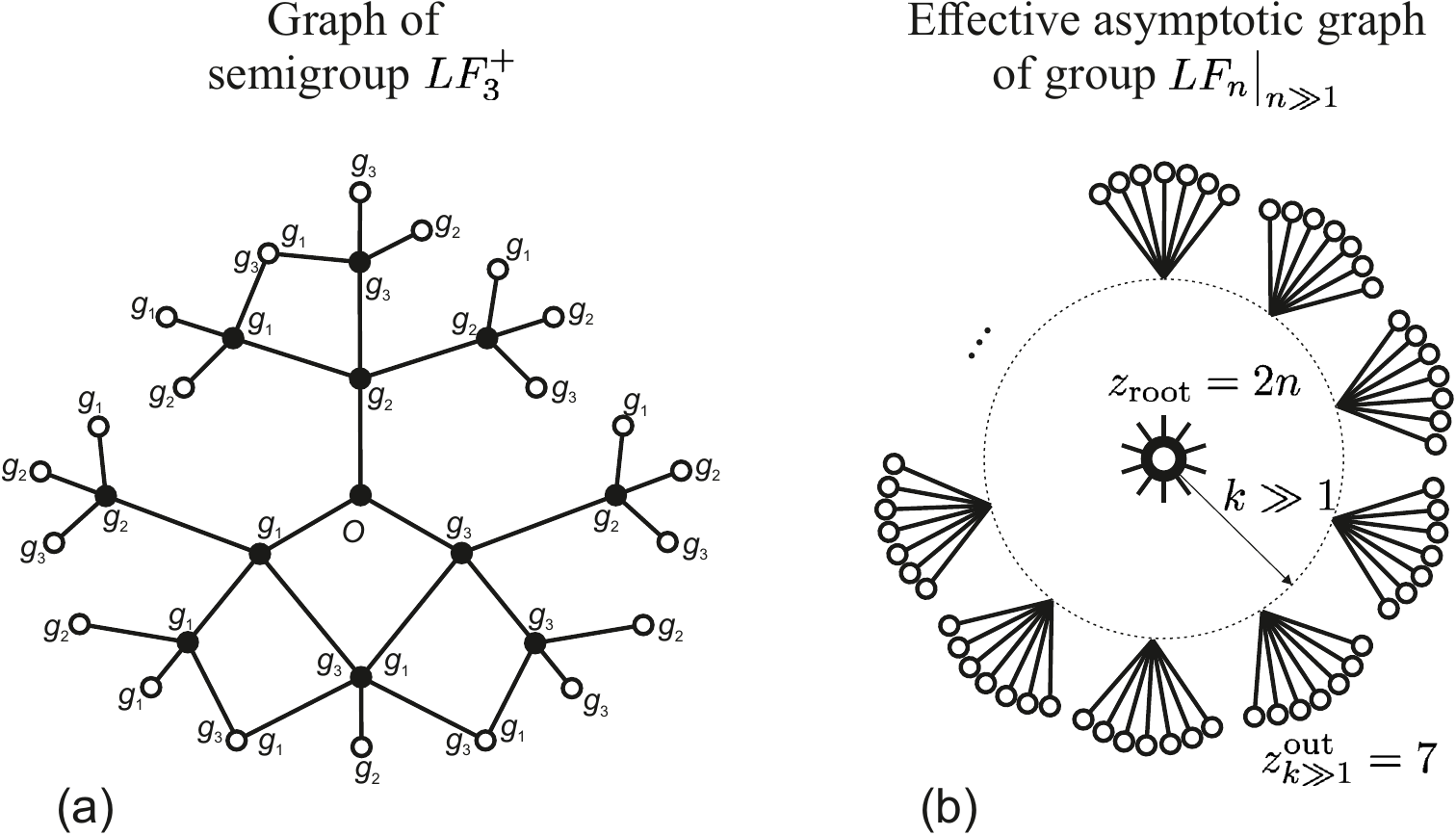}
    \caption{(a) Structure of the Cayley graph of the semigroup $LF_3^+$. (b) Effective large-scale asymptotic Cayley graph of $LF_n$ represented as a Bethe lattice with a heavy root ($z_{\rm root}=2n$, $z_{\rm out}=8$).}
    \label{fig:graph-lfn}
\end{figure}

A central challenge in this analysis stems from the complex local geometry of the graph itself, which arises from the presence of numerous shortcuts of arbitrary length. The structural complexity of the Cayley graph of $LF_n$ is already evident in the simpler case of the positive semigroup $LF_3^+$, constructed using only the set of positive generators $\{g_1, g_2, g_3\}$ of $LF_3$ -- see \fig{fig:graph-lfn}a.

To overcome the mathematical difficulties posed by this local complexity, we shift our focus to the large-scale geometric properties of $LF_n$. By analyzing the asymptotic structure of its Cayley graph at large distances from the root point, we coarse-grain short-scale loops and obtain a more tractable macroscopic limit. Thus, our approach relies on studying Brownian bridges directly on this asymptotic Cayley graph, which provides a natural framework for deriving transparent statistical results.

Each vertex of the Cayley graph corresponds to an element of $LF_n$, expressed as a word in the generators. The geodesic distance from the group identity (the root) to a given vertex equals the length of the shortest word representing that element. The asymptotic structure of the Cayley graph is governed by the growth rate of the number of non-equivalent words of a given total minimal length $k$---that is, the "volume" of the group. By computing this asymptotic volume growth, $V_{LF_n}(k)$, we characterize the asymptotic behavior of the Cayley graph of $LF_n$. It is worth emphasizing that the word length $k$ and the braid width $n$ satisfy the hierarchy $1 \ll n \ll k$. Thus, in evaluating the asymptotic volume, we first take the limit $k \to \infty$ followed by $n \to \infty$. The exponential growth of this volume at large distances $k$ is governed by the spectral radius $\rho$:
\be
\rho = \lim_{n\to\infty} \left(\lim_{k\to\infty} \left[ V_{LF_n}(k) \right]^{1/k}\right)
\label{eq:spectral_radius}
\ee
where $\rho_n = \lim_{k\to\infty} [V_{LF_n}(k)]^{1/k}$ characterizes the asymptotic exponential growth rate of shell volume, $V_{LF_n}(k) \sim \rho_n^k$, for a fixed number of generators $n$. Geometrically, the spectral radius plays the role of an effective branching factor of the Cayley graph, setting the factor by which the surface area (i.e. the volume shell) expands when passing from $k$ to $k+1$. Mathematically, $\rho_n = s_c^{-1}(n)$ corresponds to the reciprocal of the radius of convergence (critical fugacity $s_c$) of the grand canonical partition function, $Z_n(s) = \sum_{k=0}^\infty V_{LF_n}(k) s^k$, determined by the dominant eigenvalue of the transfer matrix, which sets allowed generator sequences under normal ordering.

It is worth emphasizing that the word length, $k$, and the braid width, $n$, satisfy the hierarchy $1 \ll n \ll k$. Thus, in \eq{eq:spectral_radius}, we first take the thermodynamic limit of long words, $k \to \infty$, followed by the limit of a large number of generators, $n \to \infty$.

The locally free group $LF_n$ is defined by $n$ elementary generators $\{g_1, g_2, \dots, g_n\}$ subject to the commutation relations $[g_i, g_j] = 0$ for $|i - j| > 1$. To count unique elements in $LF_n$ without double-counting equivalent words, every element $W \in LF_n$ is written uniquely as a sequence of block powers:
\be
W = g_{\alpha_1}^{m_1} g_{\alpha_2}^{m_2} ... g_{\alpha_p}^{m_p}
\ee
where $\alpha_j \in \{1, 2, \dots, n\}$ is the generator index of block $j$, and $m_j \in \mathbb{Z} \setminus \{0\}$ is a non-zero exponent. The total length of the shortest word in terms of generators is $k = |W| = \sum_{j=1}^p |m_j|$.

To select a unique canonical representative for each equivalence class, adjacent generator transitions $g_\alpha \to g_\beta$ must obey the Mikado normal ordering rule \cite{vershik2000,dudka}: 
\begin{itemize}
\item[(i)] Moving to a generator with a larger index ($\beta > \alpha$) is always permitted. For non-commuting neighbors ($g_\alpha \to g_{\alpha+1}$), moving up is allowed, while for commuting pairs ($|\beta - \alpha| > 1$), forcing $\beta > \alpha$ requires commuting generators to appear strictly in increasing order (e.g., $g_1^m g_3^q$, never $g_3^q g_1^m$);
\item[(ii)] Moving to the immediate non-commuting generator below ($\beta = \alpha - 1$) is permitted because order matters for non-commuting operators;
\item[(iii)] Transitions to the same index ($\beta = \alpha$) are forbidden, as consecutive powers of the same generator merge into a single block $g_\alpha^m$, and transitions to an index smaller by 2 or more ($\beta < \alpha - 1$) are forbidden since commuting pairs are already fixed to appear in increasing order.
\end{itemize}

The Mikado transition rules define an $n \times n$ transfer matrix $T$, where $T_{\alpha, \beta} = 1$ if the transition $g_\alpha \to g_\beta$ is allowed, and $0$ otherwise:
\be
T_{\alpha, \beta} = \begin{cases} 1, & \text{if } \beta \ge \alpha+1 \medskip \\ 
1, & \text{if } \beta = \alpha - 1 \medskip \\ 
0, & \text{otherwise}
\end{cases}
\ee
In matrix form, $T_n$ consists of $1$ everywhere strictly above the main diagonal, $1$ on the subdiagonal immediately below the main diagonal, and $0$ elsewhere,
\be
T_n = \begin{pmatrix}
0 & 1 & 1 & 1 & \dots & 1 & 1 \\
1 & 0 & 1 & 1 & \dots & 1 & 1 \\
0 & 1 & 0 & 1 & \dots & 1 & 1 \\
0 & 0 & 1 & 0 & \dots & 1 & 1 \\
\vdots & \vdots & \vdots & \ddots & \ddots & \vdots & \vdots \\
0 & 0 & 0 & \dots & 1 & 0 & 1 \\
0 & 0 & 0 & \dots & 0 & 1 & 0
\end{pmatrix}
\ee
Each block of length $l = |m| \ge 1$ has 2 possible exponent signs ($+l$ or $-l$). To compute the effective combinatorial weight $x(s)$ of a single block, assign a fugacity $s$ to each elementary step $g_\alpha^{\pm 1}$. A block of length $l = |m|$ carries the weight $s^{|m|}$. The effective weight $x(s)$ is obtained by summing over all allowed non-zero integer exponents $m \in \mathbb{Z} \setminus \{0\}$, splitting the sum into positive powers ($m \ge 1$) and negative powers ($m \le -1$),  we have:
\be
x(s) = \sum_{m \in \mathbb{Z} \setminus \{0\}} s^{|m|} = \sum_{m=1}^\infty s^m + \sum_{m=-\infty}^{-1} s^{-m}
\ee
Since positive and negative powers of length $l$ contribute identically ($s^{|-l|} = s^l$), the summation simplifies to twice the sum over all positive block lengths $l \ge 1$
\be
x(s) = 2 \sum_{l=1}^\infty s^l
\ee
Evaluating this geometric series for $|s| < 1$ we get:
\be
\sum_{l=1}^\infty s^l = s + s^2 + s^3 + \dots = \frac{s}{1-s}
\ee
Substituting this result gives the closed-form single-block weight
\be
x(s) = \frac{2s}{1-s}
\ee
At the radius of convergence $s = s_c(n)$, setting this effective single-block weight equal to the critical matrix threshold $x_c(n) = 1/\lambda_{\max}(n)$ yields
\be
x(s_c) = x_c(n) \implies \frac{2 s_c(n)}{1 - s_c(n)} = x_c(n)
\ee

A "canonical word" consisting of $p$ sequential blocks following the Mikado rules contributes a matrix factor $\left(x(s) T_n\right)^p$. Let $V_{LF_n}(k)$ be the number of distinct (non-equivalent) canonical words in the locally free group $LF_n$ that have total length $k$ (the "volume" of the group). The exponential growth rate of $V_{LF_n}$ determines the boundary growth rate of the group
\be
\rho_n = \lim_{k \to \infty} \left[V_{LF_n}(k) \right]^{1/k}
\ee
Summing over all possible block counts $p \ge 0$, the grand canonical partition function $Z_n(s)$ for all words in $LF_n$ is
\be
Z_n(s) = \sum_{k=0}^\infty V_{LF_n}(k)\, s^k = \mathbf{1}^\top \left(\sum_{p=0}^\infty \left(x(s) T_n\right)^p \right) \mathbf{1} = \mathbf{1}^\top \left(I - x(s) T_n \right)^{-1} \mathbf{1}
\ee
where $\mathbf{1} = (1, 1, \dots, 1)^\top$. The radius of convergence of $Z_n(s)$ is governed by the zeroes of the characteristic polynomial $d_n(x) = \det(I - x T_n)$.

Evaluating $d_n(x) = \det(I - x T_n)$ via row expansion yields a linear second-order recurrence relation:
\be
d_n(x) = (1 - x) d_{n-1}(x) - x^2 d_{n-2}(x)
\ee
with initial conditions $d_0(x) = 1$ and $d_1(x) = 1$. The characteristic equation associated with the recurrence, $r^2 - (1-x)r + x^2 = 0$, has roots:
\be
r_{1,2} = \frac{(1-x) \pm \sqrt{1 - 2x - 3x^2}}{2}
\ee
For $x > 1/3$, the discriminant $\Delta(x) = 1 - 2x - 3x^2$ is negative, giving complex conjugate roots $r_{1,2} = x e^{\pm i \phi}$, where the phase angle $\phi$ is defined by $\cos \phi = \frac{1 - x}{2x}$. In terms of $\phi$, $d_n(x)$ reduces to the closed form:
\be
d_n(x) = x^n \frac{\sin((n+1)\phi)}{\sin \phi}
\ee
The zeroes of $d_n(x)$ occur when $\sin((n+1)\phi) = 0$, corresponding to $(n+1)\phi_m = m\pi$. The fundamental root determining the maximal eigenvalue $\lambda_{\max}(n) = 1/x_c(n)$ corresponds to $m = 1$
\be
\phi_1(n) = \frac{\pi}{n+1}
\ee
Substituting $\phi_1(n)$ into the expression for $\cos \phi$ yields
\be
\cos \left( \frac{\pi}{n+1} \right) = \frac{1 - x_c(n)}{2 x_c(n)} \implies \lambda_{\max}(n) = \frac{1}{x_c(n)} = 1 + 2 \cos \left( \frac{\pi}{n+1} \right)
\ee
In the large-$n$ limit ($n \to \infty$), $\phi_1(n) \to 0$, and $\cos \phi_1 \to 1$. Thus, $\lim_{n \to \infty} x_c(n) = 1/3$, which implies $\lambda_{\max}(\infty) = 3$. To relate $x_c(n)$ back to the true growth rate $\rho_n$ with respect to word length $k$, we invert the single-block weight relation $x(s_c) = x_c(n)$:
\be
\frac{2 s_c(n)}{1 - s_c(n)} = x_c(n) \implies s_c(n) = \frac{x_c(n)}{2 + x_c(n)}
\ee
The exponential growth rate $\rho_n = \lim_{k \to \infty} [V_{LF_n}(k)]^{1/k}$ is the reciprocal of the critical fugacity $s_c(n)$:
\be
\rho_n = \frac{1}{s_c(n)} = \frac{2 + x_c(n)}{x_c(n)} = \frac{2}{x_c(n)} + 1 = 2 \lambda_{\max}(n) + 1
\ee
Substituting $\lambda_{\max}(n) = 1 + 2 \cos \left( \frac{\pi}{n+1} \right)$ into $\rho_n$ gives
\be
\rho_n = 2 \left[ 1 + 2 \cos \left( \frac{\pi}{n+1} \right) \right] + 1 = 3 + 4 \cos \left( \frac{\pi}{n+1} \right)
\ee
Expanding $\cos y \approx 1 - \frac{y^2}{2}$ for $y = \frac{\pi}{n+1} \approx \frac{\pi}{n}$ yields the large-$n$ asymptotic behavior
\be
\rho_n \approx 3 + 4 \left( 1 - \frac{\pi^2}{2n^2} \right) = 7 - \frac{2\pi^2}{n^2}
\ee
In the limit $n \to \infty$, the asymptotic growth rate of non-equivalent canonical words in locally free groups $LF_n$ saturates at \cite{vershik2000}
\be
\rho_\infty = \lim_{n \to \infty} \rho_n = 7
\ee

To compare this asymptotic behavior with the growth of a standard Cayley tree \footnote{Do not mix Cayley graph of the group, which reflects the topology of connections between group elements, with the Cayley tree (or Bethe lattice), which is the tree-like graph.} recall that on a regular tree $G_z$ with the coordination number $z$, the number of vertices at distance $k$ from the root is given by $V_{G_z}(k) = z(z-1)^{k-1}$ for $k \ge 1$. The exponential growth rate of such a tree is determined by its branching factor $\rho_{\text{tree}}$, where
\be
\rho_{\text{tree}} = \lim_{k \to \infty} [V_{T_z}(k)]^{1/k} = z - 1
\ee
A standard Cayley tree $G_z$ with growth rate $\rho_{\text{tree}} = 7$ corresponds to a constant vertex degree $z = 8$ uniformly for the whole graph.

In contrast, the Cayley graph of $LF_n$ exhibits a non-uniform, spatially structured branching behavior due to the commutation relations causing loops of various sizes. Locally, near the identity element ($k=1$), a step from the root can be taken along any of the $2n$ generators and their inverses, yielding a coordination number $2n$. However, as $k$ increases, the commutation relations in the group effectively restrict the effective branching factor. Taking $n \to \infty$ reveals that even with an infinite number of generators, non-commutative constraints restrict the global boundary growth rate so that it cannot exceed $\rho_\infty = \rho_{\text{tree}} = 7$. This structure is schematically depicted in \fig{fig:graph-lfn}b.

Thus, we arrive at the following schematic structure of the Cayley graph of $LF_n$: at large distances for large $n$ the growth $LF_n$ saturates and is majorated by the Cayley tree with 7 outer branches. But at the root point there are $2n$ vertices. So, the Cayley graph of $LF_n$ has the asymptotic structure of a "tree with a heavy root". In the next section we discuss the statistics of Brownian bridges on heavy-rooted trees.

\subsubsection{Localization of Brownian bridges on a Cayley tree with a heavy root}

\begin{figure}[ht]
    \centering
    \includegraphics[width=0.3\linewidth]{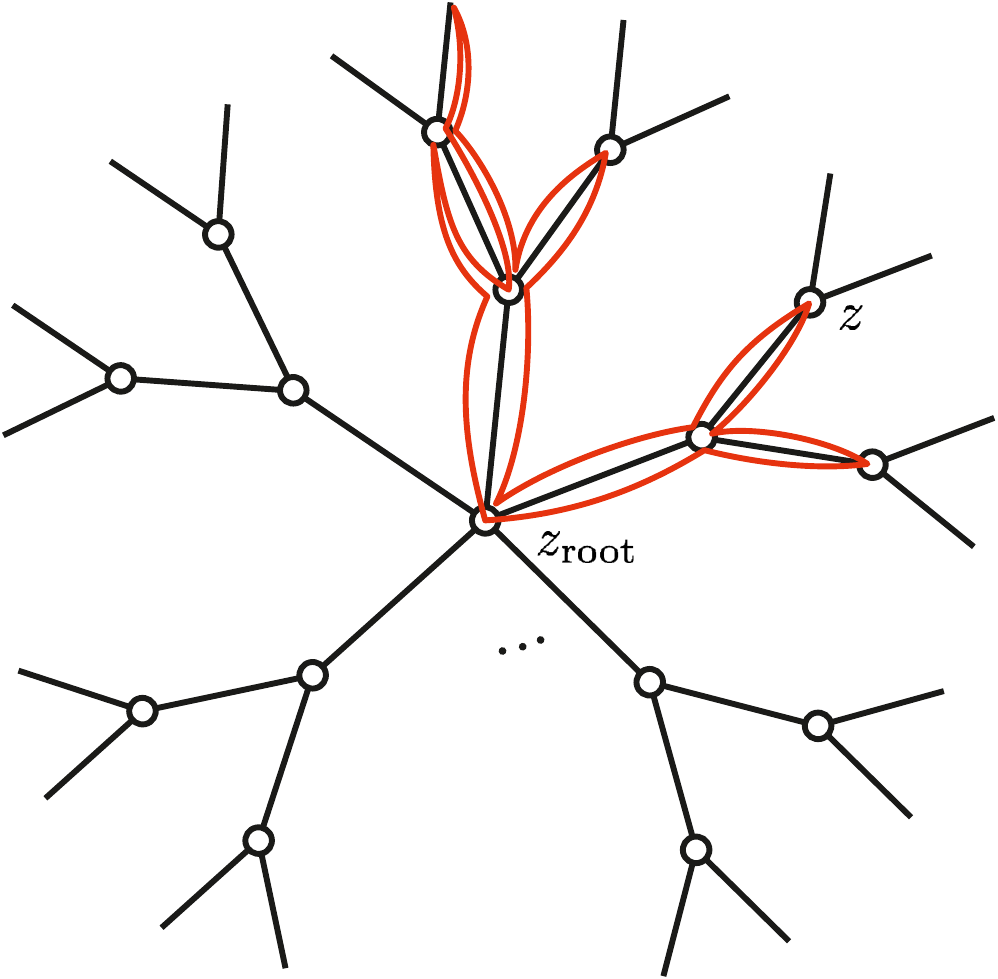}
    \caption{A heavy-rooted Cayley tree with the branching degree $z_{\rm root}$ at the root point and with a bulk branching vertex degree $z$.}
    \label{fig:heavy-root}
\end{figure}

Consider an infinite Cayley tree with a heavy root at $x=0$ having branching degree $z_{\rm root}$, while all other vertices have branching degree $z$ as shown in \fig{fig:heavy-root}. The grand partition function $W(\sigma|z,z_{\rm root})$ of open $N$-step paths originating at the root was computed in \cite{valba-tamm}. Simplifying the expression (36) of \cite{valba-tamm} we get 
\be
W(\sigma|z, z_{\rm root}) = \frac{2 z_{\rm root} \sigma \left( 1 - 2\sigma + \sqrt{1 - 4\sigma^2(z-1)} \right)}{(1 - \sigma z) \left[ 2(z-1) - z_{\rm root}\left(1 - \sqrt{1 - 4\sigma^2(z-1)}\right) \right]}
\label{eq:z-heavy}
\ee
where $\sigma$ is the fugacity of $N$, i.e. $W(\sigma|z,z_{\rm root}) = \sum_{N=0}^{\infty} W_N (z,z_{\rm root})\sigma^N$. 

The asymptotic behavior of the partition function $W_N(\sigma|z,z_{\rm root})$ at $N \to \infty$ is determined by three principal singularities in the complex $\sigma$-plane -- see \cite{valba-tamm} for detail:
\begin{enumerate}
\item The branching point $\sigma_1$, which sets the edge of the continuous spectrum. The square-root singularity corresponds to the vanishing of the radical in \eq{eq:z-heavy}:
\be
1 - 4\sigma^2(z-1) = 0 \implies \sigma_1 = \frac{1}{2\sqrt{z-1}}
\ee

\item The pole $\sigma_2$, which sets the boundary of delocalized states. Nullifying the first term in the denominator of \eq{eq:z-heavy}, we get
\be
1 - \sigma z = 0 \implies \sigma_2 = \frac{1}{z}
\ee

\item The localized pole $\sigma_3$, which sets the bound state. Nullifying the second term in the denominator of \eq{eq:z-heavy}, we obtain
\be
2(z-1) - z_{\rm root}\left(1 - \sqrt{1 - 4\sigma^2(z-1)}\right) = 0 \implies \sigma_3 = \frac{\sqrt{z_{\rm root}-z+1}}{z_{\rm root}}
\ee
\end{enumerate}

Each singularity governs a particular physical mode of path propagation encoded in the spectral features of the underlying graph adjacency matrix:

\begin{enumerate}
\item Continuous spectrum branching point $\sigma_1 = \frac{1}{2\sqrt{z-1}}$. This square-root branching point defines the outer edge of the continuous spectral band of the infinite unperturbed regular tree. Physically, it represents a delocalized, diffusive wavepacket spreading throughout the bulk. The non-analytic square-root behavior near $\sigma_1$ characterizes the algebraic decay $N^{-3/2}$ of the return probability for bulk-dominated random walks.

\item Pole $\sigma_2 = \frac{1}{z}$. This pole corresponds to a delocalized wavepacket propagating away from the root into the tree volume with a constant radial drift velocity $v = \frac{z-2}{z}$. It reflects the driving force pulling open trajectories toward infinity due to the exponential volume growth of the graph.

\item Localized bound state pole $\sigma_3 = \frac{\sqrt{z_{\rm root}-z+1}}{z_{\rm root}}$. This pole represents a discrete, exponentially localized bound state created by the entropic trap at the heavy root. When $\sigma_3$ dominates the asymptotic spectrum, the probability distribution of trajectories is pinned to $x=0$, decaying exponentially with distance $x$ from the root as $W_N(x) \sim \left(\frac{1}{z_{\rm root}-z+1}\right)^x$.
\end{enumerate}

1. For \emph{open} paths, the continuous spectrum edge $\sigma_1$ never controls the leading exponential growth because $\sigma_1 > \sigma_2$ for any regular tree with branching $z \ge 3$. The phase transition between delocalized escape and entropic localization is dictated strictly by the competition between the spreading pole $\sigma_2$ and the localized pole $\sigma_3$. Equating the two poles as in \cite{valba-tamm}, we have
\be
\sigma_2 = \sigma_3 \implies z_{\rm root} = z^2 - z
\ee
In the delocalized regime $z_{\rm root} < z^2 - z$, the pole $\sigma_2 = 1/z$ dominates. The partition function scales as $W_N \sim z^N$, and typical trajectories drift ballistically to infinity. Conversely, in the localized regime $z_{\rm root} > z^2 - z$, the bound state pole $\sigma_3$ dominates. The partition function scales as $W_N \sim \sigma_3^{-N} = \left(\frac{z_{\rm root}}{\sqrt{z_{\rm root}-z+1}}\right)^N$, and trajectories are exponentially trapped in the vicinity of the root.

2. For \emph{Brownian bridges}, all paths are constrained to start and end at the heavy root ($x_0 = x_N = 0$), and the closed-loop partition function $Z_N(0)$ counts the total number of returning paths, which manifests mathematically as the absence of the pole $\sigma_2$. As mentioned above, the pole $\sigma_2 = 1/z$ describes the probability mode carrying a drift to infinity. Imposing the global topological constraint $x_N = 0$ eliminates any trajectory that drifts away permanently. Because every path contributing to $Z_N(0)$ must form a closed loop returning to $x=0$, the mode associated with $\sigma_2$ cannot contribute to the return amplitude and should be removed from the set of singularities. Therefore, the transition for closed loops is determined by the condition $\sigma_1 = \sigma_3$:
\be
\sigma_1 = \sigma_3 \implies z_{\rm root} = 2(z-1)
\ee
Thus, the critical branching threshold for closed paths is $z_{\rm root} = 2(z-1)$. This value is consistent with the phase transition threshold discussed in \cite{burda, tamm-gulyaev}.

To conclude, we have two regimes for the statistics of Brownian bridges:
\begin{itemize}
\item In the bulk-dominated phase $z_{\rm root} < 2(z-1)$, the statistics of returning paths are controlled by the continuous spectrum edge $\sigma_1$. The closed-loop partition function scales as $W_N \sim N^{-3/2} (2\sqrt{z-1})^N$.
\item In the localized phase $z_{\rm root} > 2(z-1)$, the statistics of returning paths are controlled by the discrete bound state $\sigma_3$. The closed-loop partition function scales as $W_N \sim \left(\frac{z_{\rm root}}{\sqrt{z_{\rm root}-z+1}}\right)^N$, and paths remain exponentially localized at $x=0$.
\end{itemize}

\subsubsection{Localization of Brownian bridges on the Cayley graph of the group $LF_n$}

We now unite the results of the two previous sections to describe the statistical behavior of topological Brownian bridges on the locally free group $LF_n$. As established above, the asymptotic large-scale structure of the Cayley graph of $LF_n$ can be mapped onto a Cayley tree (Bethe lattice) with a ``heavy root''. This mapping allows us to directly apply the concept of entropic localization of Brownian bridges on a Bethe lattice to the topological Brownian bridges on the graph of $LF_n$.

For the group $LF_n$, the root vertex corresponds to the group identity, from which $2n$ branches originate, setting the local coordination number to $z_{\rm root} = 2n$. Meanwhile, the effective bulk coordination number $z$ is governed by the asymptotic boundary growth rate $\rho_n$, which we have shown saturates at $z_{\rm out}=\rho_\infty = 7$ for large $n$. This corresponds to an effective bulk coordination number $z= \rho_\infty + 1 = 8$. Recall that for Brownian bridges (closed paths), the entropic localization occurs when the heavy root degree exceeds the critical threshold $z_{\rm root} > 2(z-1)$. 

Since at $n \gg 1$ we can always reach the condition $z_{\rm root} > 2(z-1)$, and Brownian bridges on a Cayley graph of $LF_n$ are always localized in the vicinity of the root. In this localized phase, the statistical weight of the returning paths is dominated by the discrete bound state, and the paths are exponentially trapped near the group identity signaling also the spatial localization.

Whether there is a phase transition in $n$ at small $n$ ($n \sim 2 \div 5$) is yet an open question. Based on exact results for $n=2$ (which corresponds to an uniform Cayley tree with 4 branches everywhere (including the root point), and numerical results, we could expect that there is a transition from delocalized (at small $n$) to localized (at large $n$) behavior. Specifically, for $n=2$, one has $z=4$ everywhere, and since $z_{\rm root}<2(z-1) $, the system is in the bulk-dominated, delocalized regime controlled by the continuous spectrum edge. This transition in $n$ is driven by the competition between the local entropic trapping near the root point and the exponential volume growth in the bulk.

A topological Brownian bridge on the Cayley graph of $LF_n$ captures the collective behavior of all threads in the bundle, as the Brownian bridge condition requires that all threads are simultaneously topologically trivial. Since a long topological Brownian bridge ($N\gg 1$) is localized at least for $n \gg 1$, the spatial transversal span $\la |x_N|\ra $ of any thread in the bundle ($1\ll n \ll N$) remains strictly finite, $\la |x_N|\ra \sim \mathrm{const}$. Consequently, all threads are simultaneously localized and fluctuate within a tube of finite width.

\section{Conclusion and perspectives}
\label{sect:05}

\subsection{Physical mechanism: entropic traps and self-induced localization}

The statistical mechanics of unentangled directed polymer melts restricted to the trivial topological sector is governed by a self-consistent feedback loop between local transverse fluctuations and global topological accessibility. In contrast to conventional complex fluids where spatial organization is driven by energetic monomer-monomer interactions or quenched external disorder, a bundle of non-crossing directed threads experiences purely topologically induced localization.

Physically, transverse moves of individual threads alter the local topological accessibility of neighboring strands. Crucially, in a melt maintained at constant spatial density, an "entropic trap" does not correspond to a physical void; instead, it represents a preference for specific local topological configurations within the whole ensemble. These favorable configurations act as effective traps, at which unentangled chains undergo spontaneous localization. This localization launches a cascade of avalanches in the melt: as an individual strand becomes localized within an entropically favorable configuration, it imposes a cascade of topological instabilities throughout the whole fiber. Consequently, despite the absence of any interactions except topological ones, the unentangled melt undergoes emergent localization, wherein every thread of length $N$ becomes confined within a narrow tube of finite transverse span $|x_N| \sim \text{const}$ in the limit $N \to \infty$.

\subsection{Algebraic description: braid groups, locally free groups, and heavy-rooted trees}

Algebraically, the statistics of random entanglements within a bundle of $n + 1$ threads map directly onto a random walk on the Artin braid group $B_{n+1}$, or its bilateral locally-free group approximation $LF_n$. Within this group-theoretic framework, the physical condition that all threads remain globally unentangled and form unknotted loops requires the associated algebraic braid word $w_N$ to reduce identically to the group identity element $e$. The ensemble of unentangled threads thus corresponds bijectively to a topological Brownian Bridge (tBb) -- a closed random walk on the Cayley graph of the group conditioned to return to the origin at the final longitudinal step $y = N$.

On large scales, the Cayley graph of the locally-free group $LF_n$ exhibits a non-uniform geometric structure. While the local coordination number at the root is large ($z_{\text{root}} = 2n$), the asymptotic exponential growth rate of non-equivalent words in the bulk saturates at $\rho_\infty = 7$ as $n \to \infty$. Thus, macroscopically, the Cayley graph behaves as a non-uniform Cayley tree (Bethe lattice) with a "heavy root" of degree $z_{\text{root}} = 2n$ and a bulk behavior with effective coordination number $z = \rho_\infty + 1 = 8$.

The heavy root creates a structural entropic hole that lowers the free energy of returning paths. Imposing the global topological constraint of a Brownian bridge ($x_0 = x_N = 0$) suppresses the drift mode, activating a localized bound state. For $n\sim 2\div 5$, the span of a thread experiences a transition from extended algebraic behavior, typical for a homogeneous lattice of obstacles, to a localized behavior.

\subsection{Extension to (2+1)D}

Extending the considered model from (1+1)D to (2+1)D fibers offers rich theoretical perspectives for investigating the statistics and dynamics of directed random walks of diverse physical origins.

\subsubsection{Biophysics: chromosomal territories and crumpled globule}

The obtained results provide theoretical grounds for advancing our understanding of the structural stability and boundary roughness of chromosomal territories. In the context of genomics, interphase chromosomes are reasonably well-described by the crumpled (fractal) globule model \cite{grosberg1993, lieberman2009}. Due to topological uncrossability, chromatin fibers remain globally unentangled over long genomic distances. A fundamental challenge in genome organization is the observation that adjacent chromosomal territories \cite{cremer2001,cremer2010} preserve distinct spatial domains with small mutual interpenetration, rather than swelling and intermixing like standard Gaussian polymer chains.

The mechanism of collective topological localization established here suggests that unentangled chromatin strands at territorial interfaces do not diffuse freely across boundaries. Instead, self-induced entropic traps generated by neighboring unentangled loops restrict interfacial chromatin strands to narrow, self-confined tubes. This spontaneous localization acts as a robust topological barrier that suppresses large-scale transverse excursions. The roughness of the boundary separating neighboring chromosomal territories is bounded by the topological localization length.

\subsubsection{Physics: Vortex line dynamics}

A (2+1)D braid offers a space-time representation of directed linear topological defects such as vortex lines in quantum fluids or flux lines in Type-II superconductors \cite{nelson2002}. In a (2+1)D space-time representation, the trajectories of point vortices evolving in 2D can be viewed as non-crossing world-lines along the time axis $t$, forming a (2+1)D braid. One can conjecture that an ensemble of unentangled vortex lines would exhibit topological localization, wherein individual vortex lines fluctuate into localized "world-tubes", suppressing spatial vortex diffusion. This self-induced localization mechanism could be relevant to vortex clustering in thin superconducting films \cite{cordoba}.

\begin{acknowledgments}
We would like to thank G. Bunin and A. Nahum for bringing this model to our attention and for many helpful discussions.
\end{acknowledgments}

\bibliography{references.bib}

\end{document}